\documentclass[]{aa}
\usepackage{txfonts,epsfig}
\begin{document}
\title{The VLT-FLAMES Tarantula Survey. IX. The interstellar medium seen
through Diffuse Interstellar Bands and neutral sodium}
\author{Jacco Th.\ van Loon\inst{1}
\and    Mandy Bailey\inst{1}
\and    Benjamin L.\ Tatton\inst{1}
\and    Jes\'us Ma\'{\i}z Apell\'aniz\inst{2}
\and    Paul A.\ Crowther\inst{3}
\and    Alex de Koter\inst{4}
\and    Christopher J.\ Evans\inst{5}
\and    Vincent H\'enault-Brunet\inst{6}
\and    Ian D.\ Howarth\inst{7}
\and    Philipp Richter\inst{8}
\and    Hugues Sana\inst{4}
\and    Sergio Sim\'on-D\'{\i}az\inst{9,10}
\and    William Taylor\inst{5}
\and    Nolan R.\ Walborn\inst{11}
}
%\offprints{J.Th.\ van Loon}
\institute{Astrophysics Group, Lennard-Jones Laboratories, Keele University,
           Staffordshire ST5 5BG, UK
\and       Instituto de Astrof\'{\i}sica de Andaluc\'{\i}a-CSIC, Glorieta de
           la Astronom\'{\i}a s/n, 18008 Granada, Spain
\and       Department of Physics and Astronomy, University of Sheffield, Hicks
           Building, Hounsfield Road, Sheffield S3 7RH, UK
\and       Astronomical Institute Anton Pannekoek, University of Amsterdam,
           P.O.Box 94249, 1090 GE Amsterdam, The Netherlands
\and       UK Astronomy Technology Centre, Royal Observatory Edinburgh,
           Blackford Hill, Edinburgh, EH9 3HJ, UK
\and       Scottish Universities Physics Alliance (SUPA), Institute for
           Astronomy, University of Edinburgh, Royal Observatory Edinburgh,
           Blackford Hill, Edinburgh, EH9 3HJ, UK
\and       Department of Physics and Astronomy, University College London,
           Gower Street, London WC1E 6BT, UK
\and       Institut f\"ur Physik und Astronomie, Universit\"at Potsdam, Haus
           28, Karl-Liebknecht-Stra{\ss}e 24/25, D-14476 Potsdam, Germany
\and       Instituto de Astrof\'{\i}sica de Canarias, E-38200 La Laguna,
           Tenerife, Spain
\and       Departamento de Astrof\'{\i}sica, Universidad de La Laguna, E-38205
           La Laguna, Tenerife, Spain
\and       Space Telescope Science Institute, 3700 San Martin Drive,
           Baltimore, MD, 21218, USA}
\date{Submitted: August 2012}
\abstract
% context heading (optional)
{The Tarantula Nebula (a.k.a.\ 30\,Dor) is a spectacular star-forming region
in the Large Magellanic Cloud (LMC), seen through gas in the Galactic Disc and
Halo. Diffuse Interstellar Bands (DIBs) offer a unique probe of the diffuse,
cool--warm gas in these regions.}
% aims heading (mandatory)
{The aim is to use DIBs as diagnostics of the local interstellar conditions,
whilst at the same time deriving properties of the yet-unknown carriers of
these enigmatic spectral features.}
% methods heading (mandatory)
{Spectra of over 800 early-type stars from the Very Large Telescope Flames
Tarantula Survey (VFTS) were analysed. Maps were created, separately, for the
Galactic and LMC absorption in the DIBs at 4428 and 6614 \AA\ and -- in a
smaller region near the central cluster R\,136 -- neutral sodium (the Na\,{\sc
i}\,D doublet); we also measured the DIBs at 5780 and 5797 \AA.}
% results heading (mandatory)
{The maps show strong 4428 and 6614 \AA\ DIBs in the quiescent cloud complex
to the south of 30\,Dor but weak absorption in the harsher environments to the
north (bubbles) and near the OB associations. The Na maps show at least five
kinematic components in the LMC and a shell-like structure surrounding R\,136,
and small-scale structure in the Milky Way. The strengths of the 4428, 5780,
5797 and 6614 \AA\ DIBs are correlated, also with Na absorption and visual
extinction. The strong 4428 \AA\ DIB is present already at low Na column
density but the 6614, 5780 and 5797 \AA\ DIBs start to be detectable at
subsequently larger Na column densities.}
% conclusions heading (optional)
{The carriers of the 4428, 6614, 5780 and 5797 \AA\ DIBs are increasingly
prone to removal from irradiated gas. The relative strength of the 5780 and
5797 \AA\ DIBs clearly confirm the Tarantula Nebula as well as Galactic
high-latitude gas to represent a harsh radiation environment. The resilience
of the 4428 \AA\ DIB suggests its carrier is large, compact and neutral.
Structure is detected in the distribution of cool--warm gas on scales between
one and $>100$ pc in the LMC and as little as $0.01$ pc in the Sun's vicinity.
Stellar winds from the central cluster R\,136 have created an expanding
shell; some infalling gas is also detected, reminiscent of a galactic
``fountain''.}
\authorrunning{van Loon et al.}
\titlerunning{ISM seen through DIBs and Na}
\maketitle
%=========================================================================== 1
\section{Introduction}

Diffuse Interstellar Bands (DIBs) were discovered more than a century ago
(Heger 1922; cf.\ Whittet 1992). Over 400 DIBs are now known, mostly in the
4000--9000 \AA\ spectral range, displaying a rich variety in strength, width
and shape (Hobbs et al.\ 2008, 2009). They can be used to trace the diffuse,
cool--warm interstellar medium (ISM) (cf.\ Wolfire et al.\ 2003). Their
correlation with interstellar reddening, $E(B-V)$, and the abundance of atoms
(e.g., H, K, Na, Ca$^+$) and simple molecules (e.g., CN, CH, CH$^+$) ranges
from excellent to poor (Merrill \& Wilson 1938; Kre{\l}owski et al.\ 1999;
Friedman et al.\ 2011); this suggests the carriers of the DIBs are different
from ordinary dust grains, atoms and simple molecules and thus offer a unique
probe of the interstellar environment. For instance, the 5780 and 5797 \AA\
DIBs display different relative strength in harsh ($\sigma$ clouds) and mild
($\zeta$ clouds) radiation fields (Kre{\l}owski \& Westerlund 1988; Cami et
al.\ 1997; Kre{\l}owski et al.\ 1999; van Loon et al.\ 2009; Vos et al.\
2011); DIBs are generally associated with the diffuse ISM as opposed to dense
molecular clouds (see, e.g., Herbig 1993).

The situation is rather embarrassing as the carriers of the DIBs remain
unknown. The strength of the DIBs are often correlated among themselves but
not always (Cami et al.\ 1997; Moutou et al.\ 1999; McCall et al.\ 2010),
which implies a diversity of carriers. The prevailing suspicion at the moment
is that the carriers are complex, carbon-based (i.e.\ ``organic'') molecular
structures, possibly Polycyclic Aromatic Hydrocarbons (PAHs -- Crawford,
Tielens \& Allamandola 1985; van der Zwet \& Allamandola 1985; L\'eger \&
D'Hendecourt 1985); they are probably ionized, perhaps protonated (Herbig
1995), and may constitute $\sim10$\% of the total PAH abundance (Cox 2011),
the latter being measured from their infrared (IR) emission bands. Perhaps
most mysterious is the total lack of DIBs in other environments, e.g.,
circumstellar envelopes or comets (see Cox 2011), where PAHs {\it are} seen.

The strongest, and one of the broadest, the 4428 \AA\ DIB was discovered by
Merrill (1934). The profile is well represented by a Lorentzian profile, which
Snow, Zukowski \& Massey (2002) interpreted as due to broadening by natural
damping in a molecular carrier; they also showed ``saturation'' of the growth
of equivalent width for $E(B-V)>1$ mag, which they explained by the carrier
being present in the skins of clouds but not deeper inside (see also Herbig
1995). The 6614 \AA\ DIB reaches similar depth as the 4428 \AA\ DIB but is
much narrower; it displays sub-structure resembling a blend of transitions
(Sarre et al.\ 1995), which was taken as further evidence for a molecular
carrier (see also Cami et al.\ 2004). These two DIBs are the main focus of the
study we present here.

DIBs have been detected in the ISM of the Magellanic Clouds (e.g., Walker
1963; Hutchings 1964; Vladilo et al.\ 1987; Ehrenfreund et al.\ 2002), whose
red-shifts of $\sim180$ km s$^{-1}$ (Small Magellanic Cloud -- SMC) and
$\sim270$ km s$^{-1}$ (Large Magellanic Cloud -- LMC) allow easy separation of
Galactic foreground absorption and internal Magellanic Cloud absorption for
relatively narrow DIBs such as the 6614 \AA\ DIB but not for the broader DIBs
such as the 4428 \AA\ DIB. Any dependency on metallicity ($\sim20$\% and
$\sim50$\% solar in the SMC and LMC, respectively), or lack thereof, could
reveal clues about the composition of the carrier of the DIB. Welty et al.\
(2006), for instance, found both the 5780 {\it and} 5797 \AA\ DIBs are weaker
by a factor 7--9 in the LMC and 20 in the SMC relative to H\,{\sc i} in
comparison to Galactic sightlines, a stronger difference than explained by
metallicity alone; individual elemental abundances may be key, such as the
nitrogen abundance which is more depleted than carbon in the Magellanic Clouds
(van Loon et al.\ 2010a,b). Cox et al.\ (2006, 2007) suggest that besides
metallicity, the generally harsher ultraviolet (UV) radiation field in
lower-metallicity ISM is the other factor determining the behaviour of the
DIBs in the Magellanic Clouds.

Multi-object or integral-field spectroscopic capabilities offer an efficient
way in which to map the ISM. Following the first atomic absorption-line map
of the intervening ISM towards the Galactic globular cluster
$\omega$\,Centauri (van Loon et al.\ 2007) we also mapped the 5780 and 5797
\AA\ DIBs in that direction and found a high ratio indicative of the harsh
radiation field that characterises the extra-planar environment of the Milky
Way (van Loon et al.\ 2009). We have since embarked on a programme to
construct maps of the DIB strength across large areas of the Magellanic
Clouds, and here present the first such map of the 4428 and 6614 \AA\ DIBs
covering the Tarantula Nebula in the LMC, arguably the most prolific
star-forming region in the Local Group, as well as Galactic high-latitude
foreground gas.

%
% TABLE 1
%
\begin{table}
\caption{Summary of the optical spectral data: VLT--FLAMES mode, setting and
spectral resolution; number of sightlines (different stars); and the DIBs
analysed by us in detail (for a brief discussion of the 4502, 4727, 4762 and
4780 \AA\ DIBs see Section 4.1).}
\begin{tabular}{llccccc}
\hline\hline
Mode                            &
setting                         &
\llap{F}WH\rlap{M}              &
\multicolumn{3}{c}{sightlines}  &
DIB                             \\
                                &
                                &
(\AA)                           &
available                       &
\multicolumn{2}{c}{accepted}    &
(\AA)                           \\
                                &
                                &
                                &
                                &
LMC                             &
Galactic                        &
                                \\
\hline
Medusa                          &
LR02                            &
0.61                            &
\llap{8}93                      &
\llap{7}86                      &
\llap{7}86                      &
4428                            \\
                                &
HR15\rlap{N}                    &
0.41                            &
\llap{8}93                      &
\llap{6}57                      &
\llap{6}08                      &
6614                            \\
ARGUS                           &
LR02                            &
0.40                            &
42\rlap{$^a$}                   &
40\rlap{$^a$}                   &
40\rlap{$^a$}                   &
4428                            \\
UVES                            &
520(L)                          &
0.10                            &
25                              &
24                              &
24                              &
4428                            \\
                                &
520(U)                          &
0.10                            &
25                              &
25                              &
24                              &
5780                            \\
                                &
520(U)                          &
0.10                            &
25                              &
25                              &
25                              &
5797                            \\
\hline
\end{tabular}

\medskip
Notes: $a$ -- includes a spectrum of the unresolved core of R\,136.
\end{table}

%
% FIGURE 1
%
\begin{figure}
\centerline{\vbox{
\psfig{figure=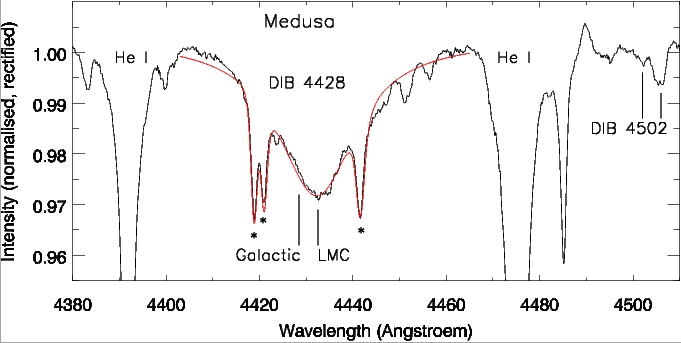,width=90mm}
\psfig{figure=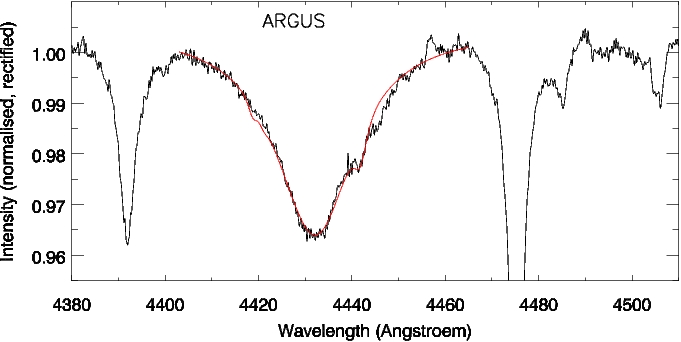,width=90mm}
\psfig{figure=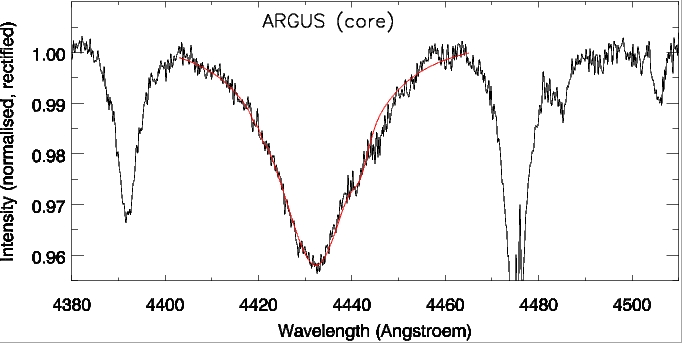,width=90mm}
\psfig{figure=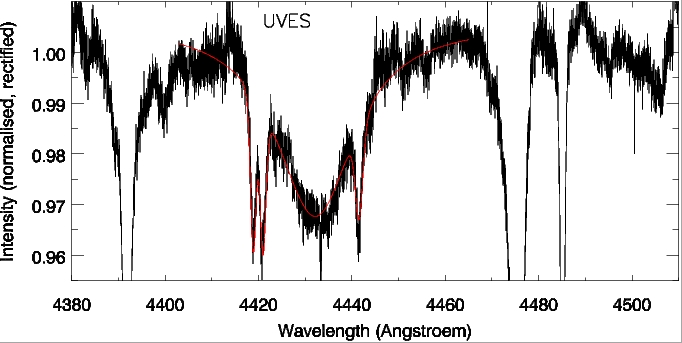,width=90mm}
}}
\caption[]{Normalised and rectified, r.m.s.(squared)-weighted average spectra
of the 4428 \AA\ DIB, with overplotted Lorentzian profiles fit to the
individual spectra, in (from {\it Top} to {\it Bottom}:) the Medusa sample;
the ARGUS sample; the ARGUS spectrum of the unresolved core of R\,136; the
UVES sample. Three sharp stellar photospheric lines that were fit
simultaneously with the DIB are marked with stars (see Appendix). Spectra were
included only if reasonable fits could be obtained.}
\label{spectra1}
\end{figure}

%
% FIGURE 2
%
\begin{figure}
\centerline{\vbox{\hbox{
\hspace{-1.5mm}
\psfig{figure=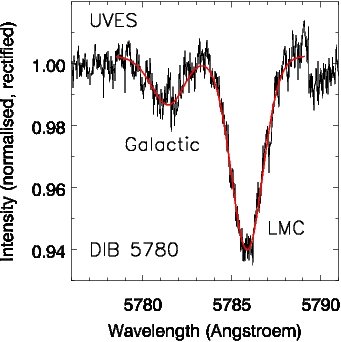,width=44.5mm}
\psfig{figure=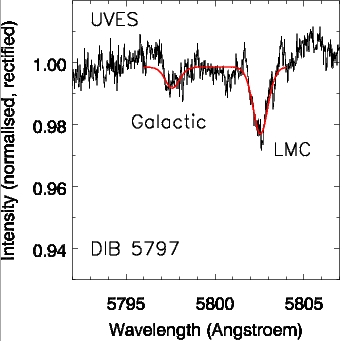,width=44.5mm}}
\psfig{figure=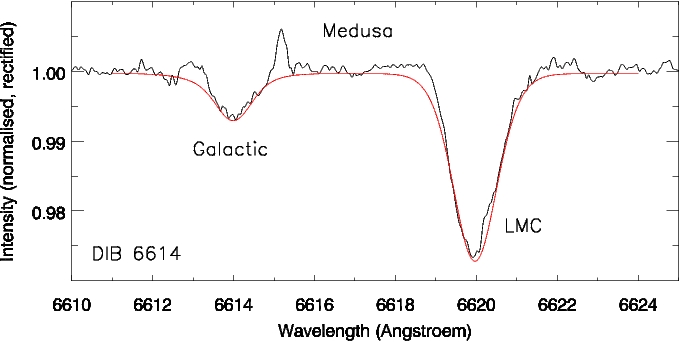,width=90mm}
}}
\caption[]{Normalised and rectified, r.m.s.(squared)-weighted average spectra
of the ({\it Top left:}) 5780 and ({\it Top right:}) 5797 \AA\ DIBs in the
UVES sample and ({\it Bottom:}) 6614 \AA\ DIB in the Medusa sample; with
overplotted Gaussian profiles fit to the individual spectra. Spectra were
included only if reasonable fits could be obtained (see Appendix).}
\label{spectra2}
\end{figure}

%=========================================================================== 2
\section{Measurements}

The optical spectra were all obtained as part of the Very Large Telescope
FLAMES Tarantula Survey (VFTS; Evans et al.\ 2011). They comprise three sets,
arising from the use of different modes and settings of the FLAMES instrument
(Pasquini et al.\ 2002), of which Table 1 summarises the spectral resolution
(Full Width at Half Maximum -- FWHM), the number of sightlines we extracted,
and the DIB that we analyse here. Additional DIBs are covered by the spectral
ranges but they are too weak for a meaningful analysis of the kind intended
here. We also searched for the sharp lines from CH and CH$^+$ molecules at
4300 and 4232 \AA, respectively (cf.\ Welty et al.\ 2006), but found no trace
of them above a 3-$\sigma$ level of $\sim0.01$ \AA; this is not unexpected as
in Welty et al.'s sample of 20 sightlines only one exceeded this limit (and
only in CH).

%------------------------------------------------------------------------- 2.1
\subsection{Diffuse Interstellar Bands}

The Medusa spectra were acquired using different settings with somewhat
different spectral resolution. A total of 800 sightlines were analysed,
comprising $\sim300$ O-type stars and $\sim500$ B-type stars; 93 additional
spectra of A- or later type stars were excluded from the present analysis as
the interstellar features are more affected by the photospheric lines. Six
spectra were obtained for each star, within a year (some stars were observed
more often). These spectra were normalised and rectified in the spectral
region surrounding the interstellar features using a $2^{\rm nd}$-order
polynomial, before being averaged using weights according to the r.m.s.\
scatter in the ``continuum'' regions chosen to be as little troubled by
spectral features as possible. The r.m.s.(squared)-weighted averages of those
(normalised and rectified) spectra with reasonable fits (see Appendix) to the
4428 and 6614 \AA\ DIBs are displayed in Figs.\ 1 and 2, respectively.

The ARGUS spectra were acquired with the same LR02 setting also used with
Medusa, on five occasions. ARGUS is an integral-field unit which was used to
more reliably extract the spectra of stars in the proximity of the R\,136
cluster core -- a spectrum of the unresolved core was also extracted. These
spectra are dominated by O-type stars, in contrast to the Medusa spectra that
are dominated by B-type stars. Stars VFTS\,542, 570 and 585 were observed both
with Medusa and ARGUS. The ARGUS spectra were treated in the same manner as
the Medusa spectra. The r.m.s.(squared)-weighted average of those (normalised
and rectified) spectra with reasonable fits (see Appendix) to the 4428 \AA\
DIB is displayed in Fig.\ 1, along with the spectrum of the unresolved core of
R\,136.

The UVES spectra were acquired at the same time as the ARGUS spectra. They
have a significantly higher resolving power as well as wider spectral range,
providing access to the 4428 \AA\ DIB as well as two other strong DIBs at 5780
\AA\ and 5797 \AA, and the atomic doublet of Na\,{\sc i}\,D at 5889.95 and
5895.92 \AA. In total, 25 sightlines were analysed, all in the inner part of
30\,Dor. All have also Medusa spectra except VFTS\,416, 482, 545, 562 and 641
(VFTS\,545 does have an ARGUS spectrum). VFTS\,542 was observed with UVES as
well as ARGUS {\it and} Medusa. The UVES spectra were treated in the same
manner as the Medusa spectra. The r.m.s.(squared)-weighted average of those
(normalised and rectified) spectra with reasonable fits (see Appendix) to the
4428 \AA\ DIB is displayed in Fig.\ 1; with the 5780 and 5797 \AA\ DIBs
displayed in Fig.\ 2.

The first thing to notice is that the 4428 \AA\ DIB absorption trough spans
$\sim60$ \AA, far wider than the separation of the Galactic and LMC components
($\Delta\lambda_{4428}\sim4$ \AA), whilst the 6614 \AA\ DIB is sufficiently
narrow for the two components to be well resolved ($\Delta\lambda_{6614}\sim6$
\AA). The second thing to notice is that the LMC component is, at least on
average for the 6614 \AA\ DIB, almost four times stronger than the Galactic
component. We also note that the 4428 \AA\ DIB towards R\,136 is stronger than
what is typical across the wider region of the VFTS. Finally, the 4502 \AA\
DIB (Herbig 1975) is clearly present in the LMC but its Galactic absorption is
very weak (Fig.\ 1; see Section 4.1).

The strongest DIBs are detected in VFTS\,232, at a peak depth of 11\% and 15\%
of the continuum for the 4428 and 6614 \AA\ DIBs, respectively (see Fig.\ A2).
The runaway star VFTS\,16 reported in Evans et al.\ (2010) was noted for its
prominent 4428 \AA\ DIB, but the DIBs in VFTS\,16 are several times weaker
than in VFTS\,232 and in fact not atypical for the region in the Tarantula
Nebula covered by the VFTS.

%------------------------------------------------------------------------- 2.2
\subsection{Sodium}

%
% FIGURE 3
%
\begin{figure}
\centerline{\psfig{figure=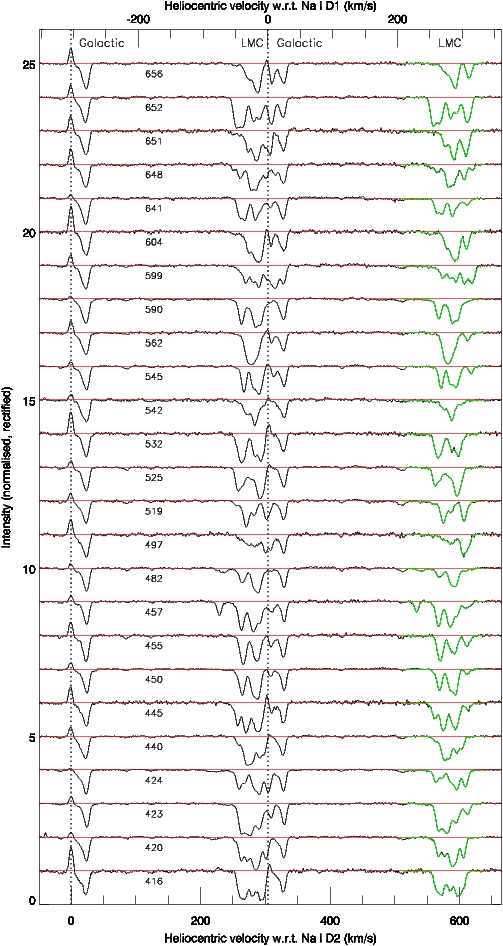,width=90mm}}
\caption[]{Normalised and rectified spectra of the Na\,{\sc i}\,D doublet in
the UVES spectra; each spectrum is offset from the one below by unity. The
vertical dotted lines mark telluric emission at the rest wavelengths.
Overplotted in green are the combined fits of ensembles of Gaussians to the
D$_1$ line centred between 5900.3--5902.2 \AA.}
\label{NaD}
\end{figure}

The Na\,{\sc i}\,D doublet of neutral sodium atoms is superbly detected in all
UVES spectra both for the Galactic and LMC component (Fig.\ 3). The D$_1$
absorption from the LMC component (5900.3--5902.2 \AA) is unaffected by other
spectral features, as judged from the correspondence with the D$_2$ absorption
(5894.2--5896.1 \AA), but the D$_2$ absorption from the LMC component is not
(always) completely separate from the D$_1$ absorption from the Galactic
component. The absorption from the Galactic component is generally affected
(if sometimes only very slightly) by sharp telluric emission at the rest
wavelength (which also affects the D$_2$ absorption from the LMC component).

The Galactic absorption has one dominant component red-shifted by $\sim0.4$
\AA\ ($\sim20$ km s$^{-1}$, i.e.\ comparable to the local standard of rest
velocity in that direction of 15 km s$^{-1}$) but also a (much) weaker
component red-shifted by about half that amount. The latter shows clear signs
of variations among the different sightlines. The LMC absorption has multiple
components -- five or more -- spanning a heliocentric velocity range of about
230--320 km s$^{-1}$; there are huge variations among the different sightlines.
While interesting (see Sections 3.3 and 4.2), this clearly complicates the
analysis of the DIBs, which are not resolved at this level of detail. We have
attempted to quantify the kinematic components in the D$_1$ line by
simultaneously fitting seven Gaussians centred between 5900.3 and 5902.2 \AA\
(Table 2; the combined fits are overplotted in Fig.\ 3). Seven was the minimum
number of Gaussians needed to satisfactorily identify all the components that
can be readily recognised by visual inspection, for every spectrum; in most
cases one or more of these Gaussian components were found to be insignificant,
and we have omitted from Table 2 those fitted components with a peak depth
below one per cent of the continuum.

The analysis of the Na\,{\sc i}\,D absorption is complicated as the strongest
components are saturated: the oscillator strengths of the D$_1$ and D$_2$
transitions compare as 1:2 but the observed ratio in our spectra is much
closer to unity.

%
% TABLE 2
%
\begin{table*}
\caption{Kinematic components in the Na\,{\sc i}\,D$_1$ absorption most likely
to arise from gas within the LMC, based on simultaneous fitting of Gaussians.
Listed are: central heliocentric velocity and FWHM (in km s$^{-1}$) and
central depth, $I_{\rm c}$ (w.r.t.\ continuum, larger values signifying deeper
absorption).}
\begin{tabular}{l|crc|crc|crc|crc|crc|crc|crc}
\hline\hline
VFTS & \multicolumn{3}{|c}{1} &
       \multicolumn{3}{|c}{2} &
       \multicolumn{3}{|c}{3} &
       \multicolumn{3}{|c}{4} &
       \multicolumn{3}{|c}{5} &
       \multicolumn{3}{|c}{6} &
       \multicolumn{3}{|c}{7} \\
 & $v_{\rm c}$ & {\tiny \llap{FW}H\rlap{M}} & $I_{\rm c}$ &
   $v_{\rm c}$ & {\tiny \llap{FW}H\rlap{M}} & $I_{\rm c}$ &
   $v_{\rm c}$ & {\tiny \llap{FW}H\rlap{M}} & $I_{\rm c}$ &
   $v_{\rm c}$ & {\tiny \llap{FW}H\rlap{M}} & $I_{\rm c}$ &
   $v_{\rm c}$ & {\tiny \llap{FW}H\rlap{M}} & $I_{\rm c}$ &
   $v_{\rm c}$ & {\tiny \llap{FW}H\rlap{M}} & $I_{\rm c}$ &
   $v_{\rm c}$ & {\tiny \llap{FW}H\rlap{M}} & $I_{\rm c}$ \\
\hline
416&   &  &   &259& 8&.41&268&11&.70&280& 8&.48&294&15&.77&303& 5&.21&   &  &   \\
420&   &  &   &262& 6&.32&272&18&.52&283& 7&.46&289& 9&.70&301& 8&.60&313& 5&.01\\
423&245&15&.04&262& 9&.65&273&13&.76&279& 8&.34&290&12&.56&306&10&.25&310& 6&.11\\
424&   &  &   &259& 9&.35&269& 7&.22&283&10&.36&292& 8&.53&304& 9&.49&   &  &   \\
440&   &  &   &259&10&.15&272&10&.68&280& 8&.55&291&10&.49&300& 7&.31&311& 5&.02\\
445&232& 5&.02&258&11&.46&268& 6&.38&273& 9&.69&283& 5&.15&289&13&.78&308& 8&.20\\
450&243& 8&.01&255& 6&.07&265& 9&.64&281&10&.61&290& 8&.67&   &  &   &307& 8&.15\\
455&232&18&.02&   &  &   &266&10&.77&281& 7&.30&288&10&.51&   &  &   &307&10&.31\\
457&229& 7&.31&263&10&.74&253& 5&.05&281&13&.77&292& 7&.31&   &  &   &306&14&.20\\
482&237&11&.07&251& 5&.02&265&12&.25&282& 9&.42&290& 9&.46&304&10&.07&315& 5&.03\\
497&239& 5&.04&254& 5&.08&272&14&.15&280& 5&.06&285&12&.23&301& 9&.60&309& 8&.35\\
519&235& 5&.03&256&10&.07&271&10&.63&282& 8&.31&290& 5&.03&301& 8&.43&305&10&.22\\
525&243&13&.02&257& 7&.35&263&11&.40&282&18&.30&291& 9&.65&298& 8&.37&313& 7&.03\\
532&238&18&.03&259& 8&.18&264&13&.64&283& 8&.47&303& 5&.08&294&10&.66&317& 9&.02\\
542&   &  &   &257&18&.04&270& 9&.22&284& 5&.12&283&12&.49&298& 9&.11&313& 5&.02\\
545&   &  &   &254& 5&.03&267& 8&.65&281& 7&.41&290&12&.61&   &  &   &313& 7&.21\\
562&   &  &   &253& 5&.05&274&12&.82&283& 9&.55&290& 7&.32&297& 5&.03&308& 9&.22\\
590&   &  &   &267&18&.16&263& 8&.43&285&11&.68&293& 7&.44&301&14&.11&   &  &   \\
599&248& 5&.03&266&18&.10&270& 7&.26&281&10&.35&290& 7&.42&303&10&.52&315& 9&.51\\
604&   &  &   &251&13&.08&276&11&.44&286&10&.76&293& 8&.55&   &  &   &306& 8&.51\\
641&223&18&.02&260& 8&.40&269& 9&.44&284& 6&.28&288&16&.29&299& 5&.01&307& 9&.18\\
648&   &  &   &261&17&.20&   &  &   &277& 7&.38&285&14&.60&303& 8&.44&315& 7&.17\\
651&232& 8&.04&254&13&.06&273& 7&.23&278&18&.30&287&10&.70&307& 9&.51&302&11&.34\\
652&   &  &   &254& 9&.76&265&12&.77&281& 8&.59&291&11&.47&306&10&.73&313& 6&.28\\
656&239& 5&.02&254& 7&.04&276&15&.30&286&10&.49&291& 7&.39&   &  &   &310& 9&.44\\
\hline
\end{tabular}
\end{table*}

We resolved to measure the equivalent width by integration of the spectrum
(i.e.\ {\it not} by fitting analytical functions), of the D$_1$ line of the
LMC component and of the D$_2$ line of the Galactic component (just avoiding
the telluric emission). As these appear to reach similar maximum depth they
will also be affected to a similar degree by saturation effects.

%=========================================================================== 3
\section{Analysis}

%------------------------------------------------------------------------- 3.1
\subsection{Maps of interstellar absorption}

%
% FIGURE 4
%
\begin{figure*}
\centerline{\vbox{\hbox{
\psfig{figure=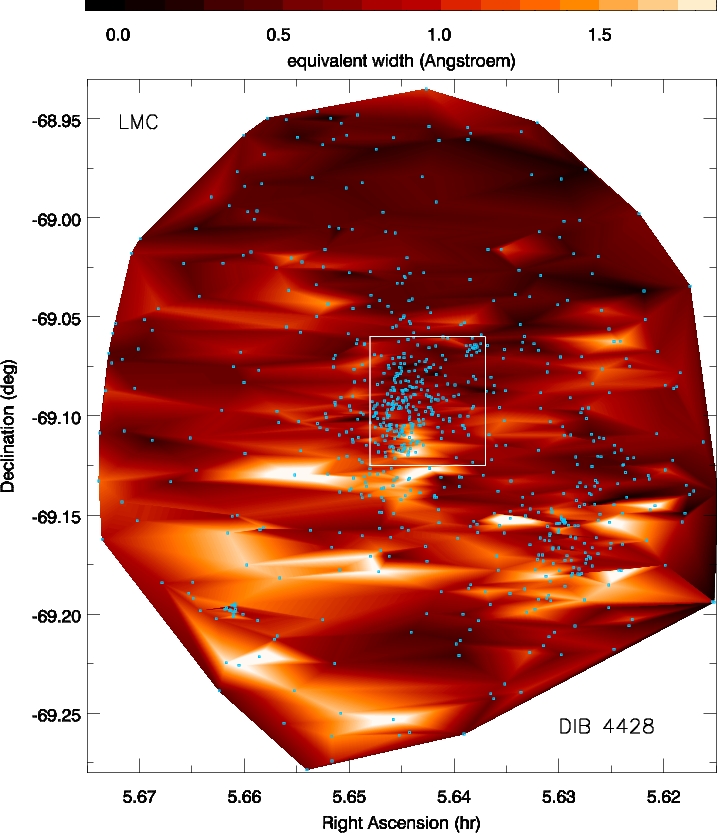,width=61mm}
\psfig{figure=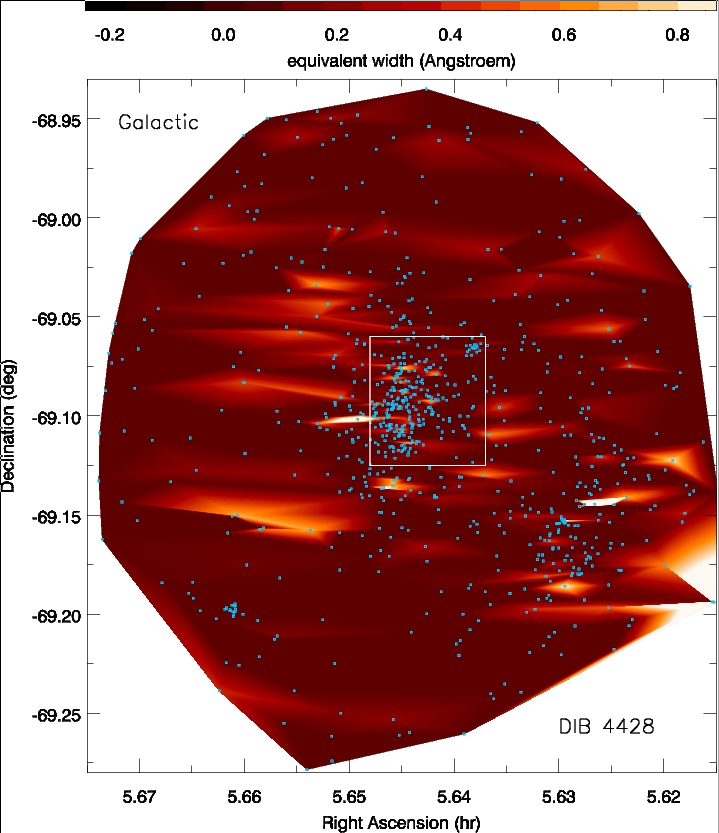,width=61mm}
\psfig{figure=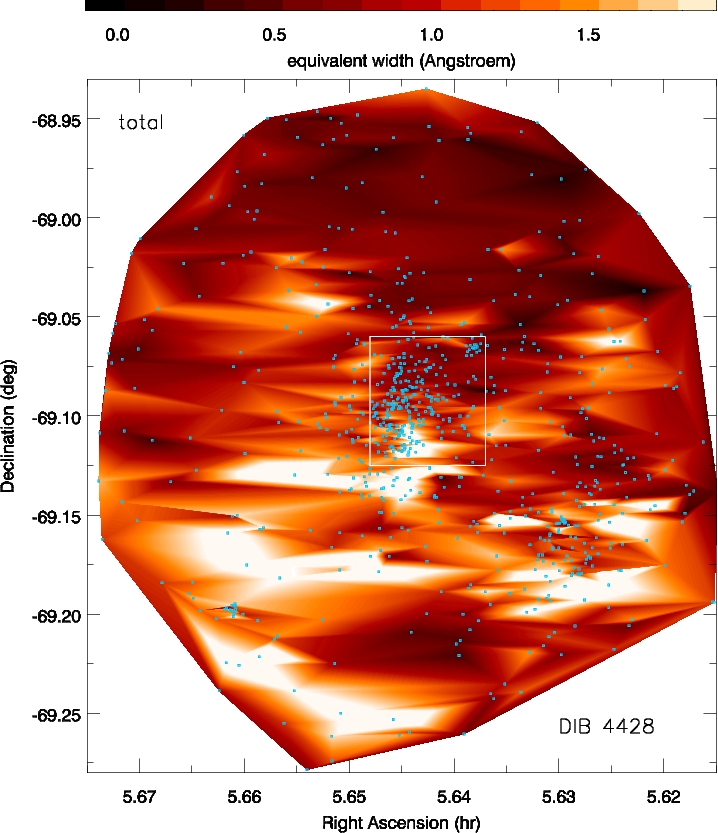,width=61mm}
}\vspace{2mm}\hbox{
\psfig{figure=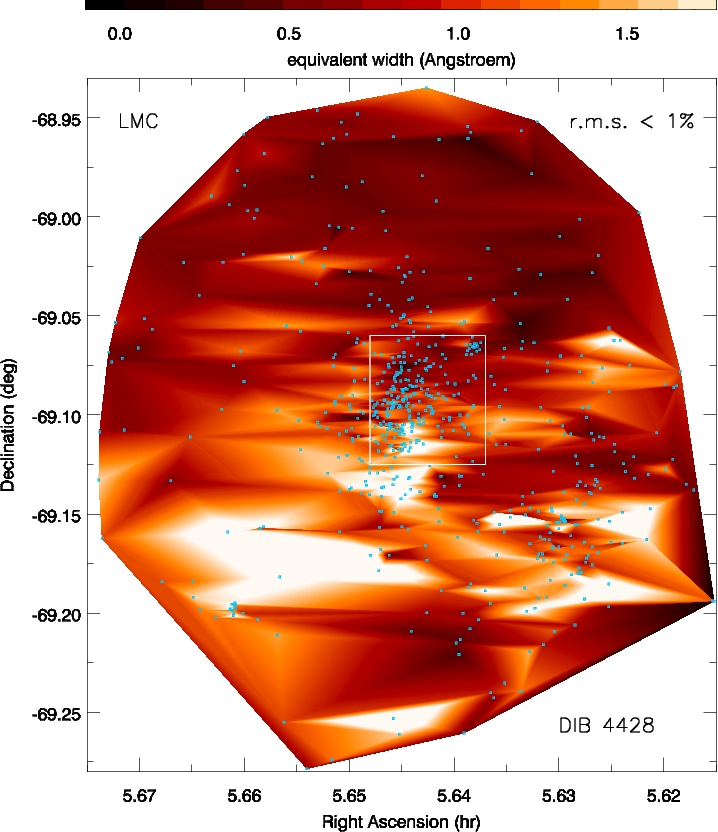,width=61mm}
\psfig{figure=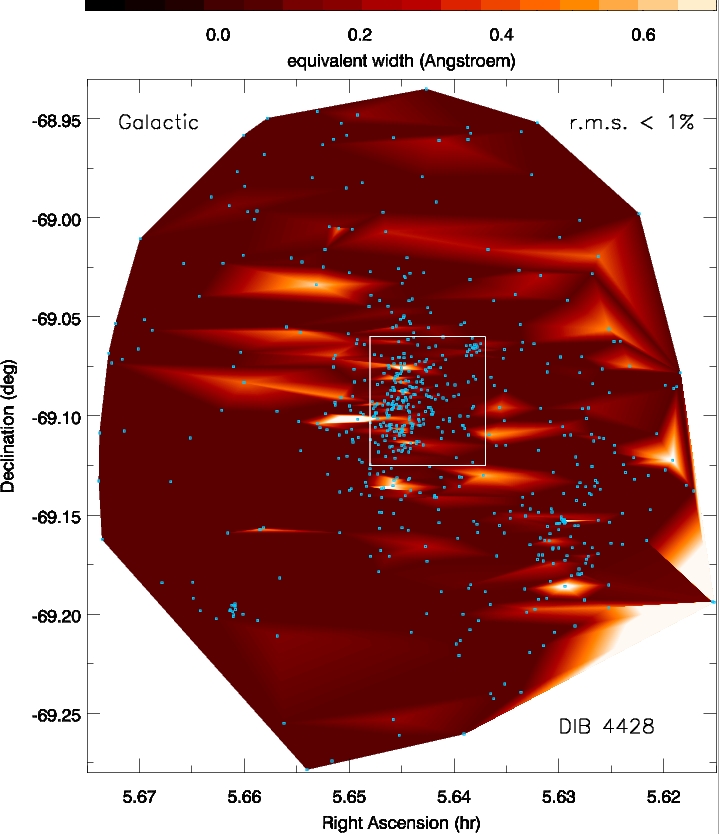,width=61mm}
\psfig{figure=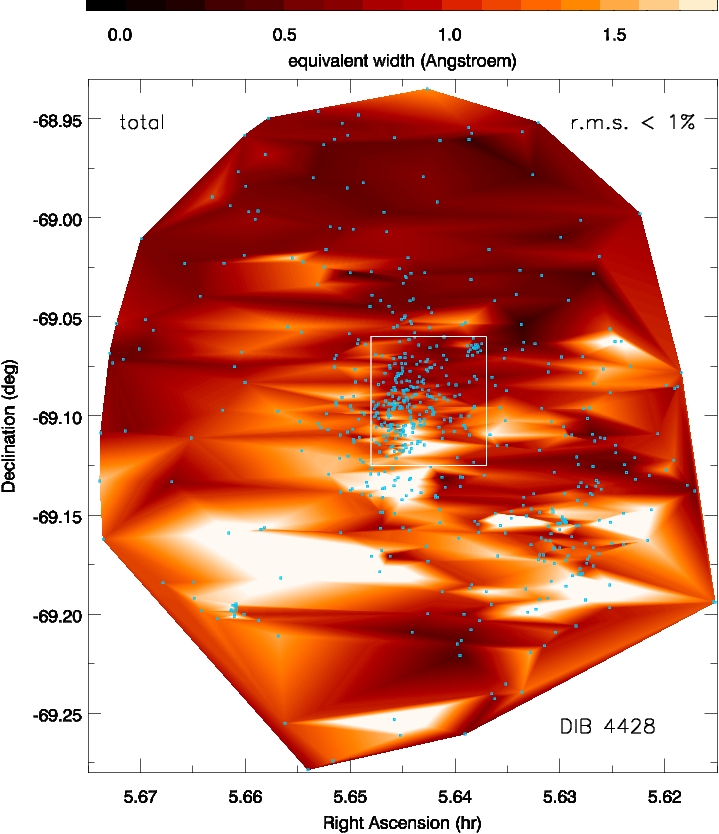,width=61mm}
}}}
\caption[]{Equivalent width maps of the 4428 \AA\ DIB absorption in the ({\it
Left:}) LMC component, ({\it Middle:}) Galactic component and ({\it Right:})
LMC + Galactic components, for ({\it Top:}) all and ({\it Bottom:}) the best
Medusa spectra where reasonable fits could be obtained. The maps cover an area
of $19^\prime({\rm RA})\times21^\prime({\rm Dec})$, which corresponds to
$0.28\times0.31$ kpc$^2$ at the distance of the LMC. The sightlines are marked
with little blue dots. The white box delineates an area which is displayed in
more detail in Fig.\ 5.}
\label{map1}
\end{figure*}

%
% FIGURE 5
%
\begin{figure*}
\centerline{\vbox{\hbox{
\psfig{figure=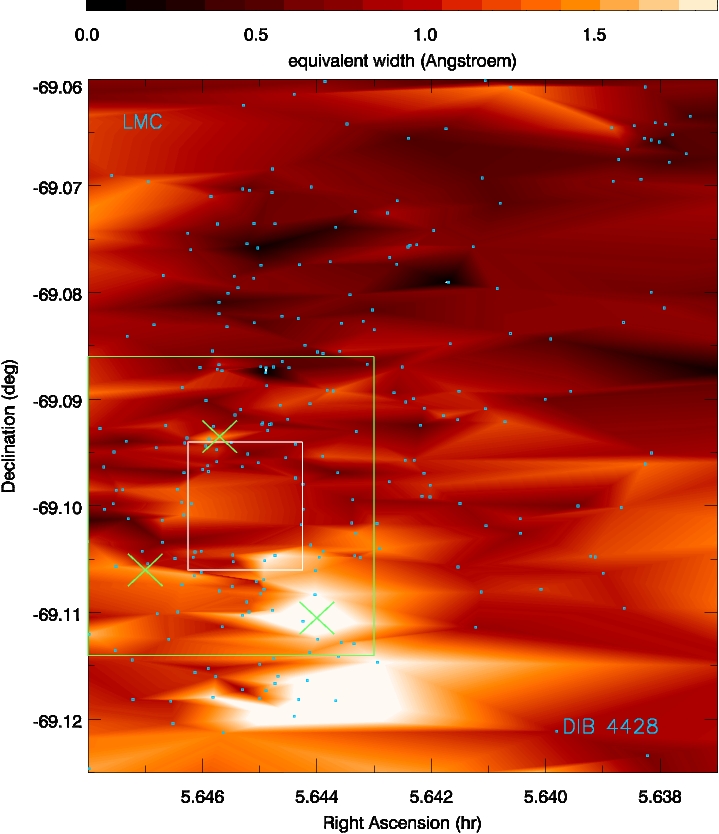,width=61mm}
\psfig{figure=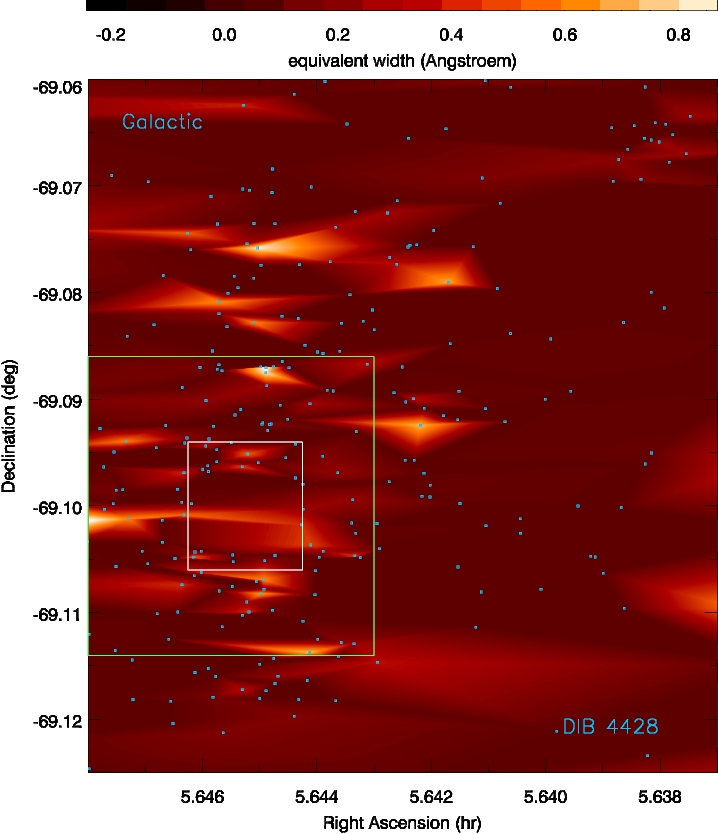,width=61mm}
\psfig{figure=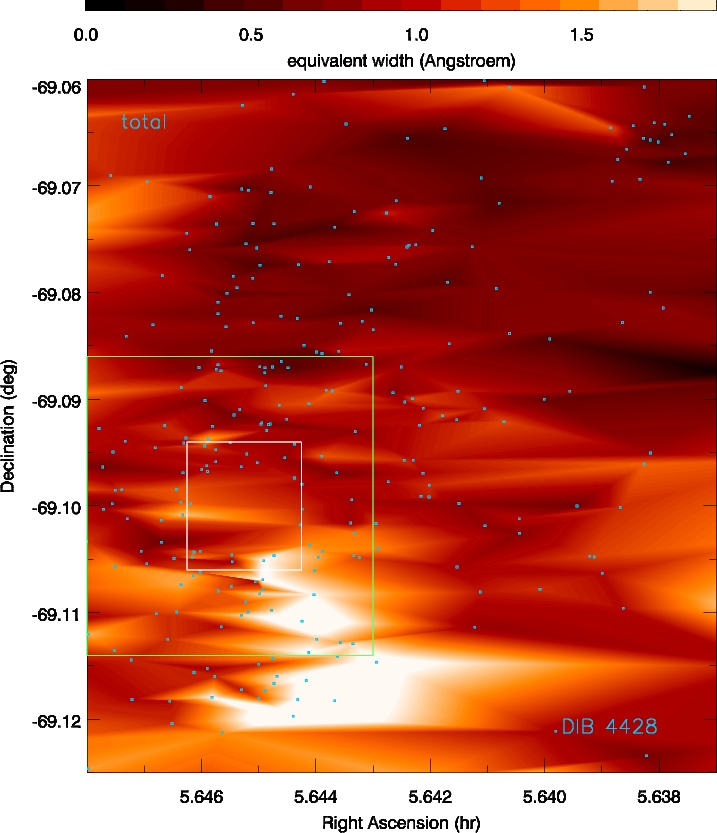,width=61mm}
}\vspace{2mm}\hbox{
\psfig{figure=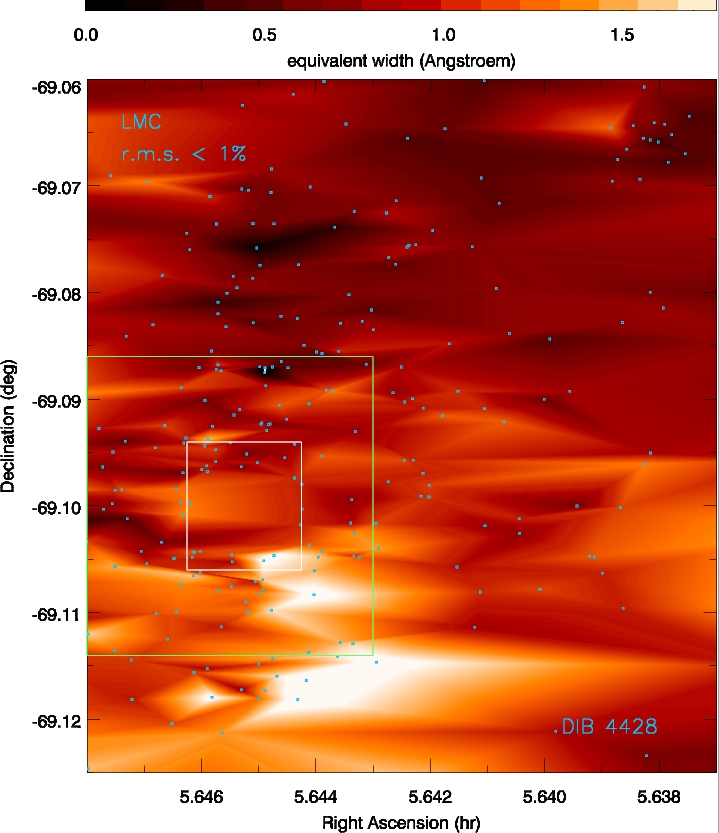,width=61mm}
\psfig{figure=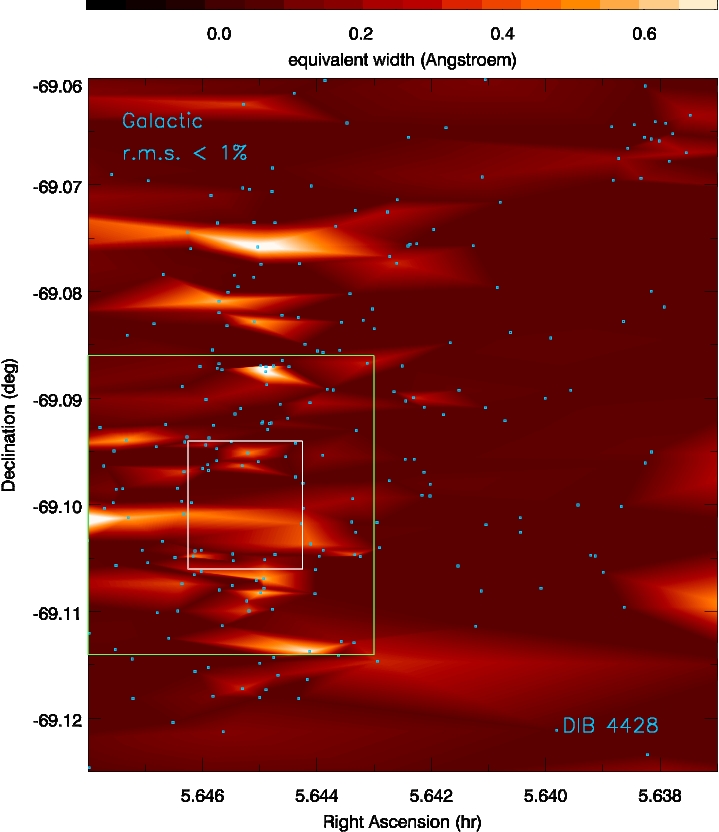,width=61mm}
\psfig{figure=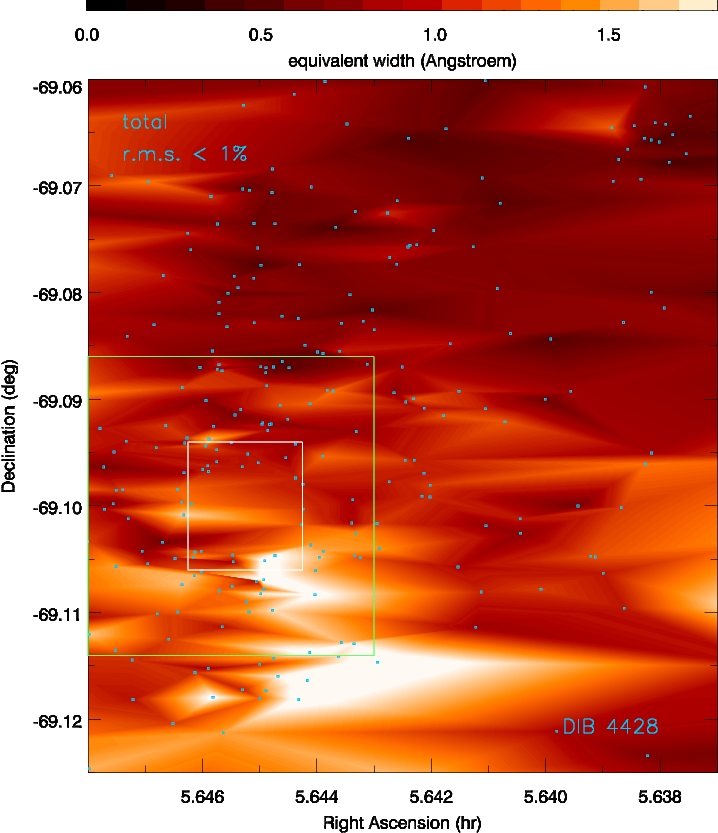,width=61mm}
}}}
\caption[]{As Fig.\ 4 but zoomed in on the densely sampled central region
measuring $3\rlap{.}^\prime5({\rm RA})\times3\rlap{.}^\prime9({\rm Dec})$,
which corresponds to $51\times57$ pc$^2$ at the distance of the LMC. The white
box delineates the area displayed in Fig.\ 8 (ARGUS spectra), while the green
box delineates the area of the Na maps (Figs.\ 12 \& 13). The green crosses
mark three LMC clouds also seen in Na absorption (Fig.\ 12).}
\label{map2}
\end{figure*}

%
% FIGURE 6
%
\begin{figure*}
\centerline{\vbox{\hbox{
\psfig{figure=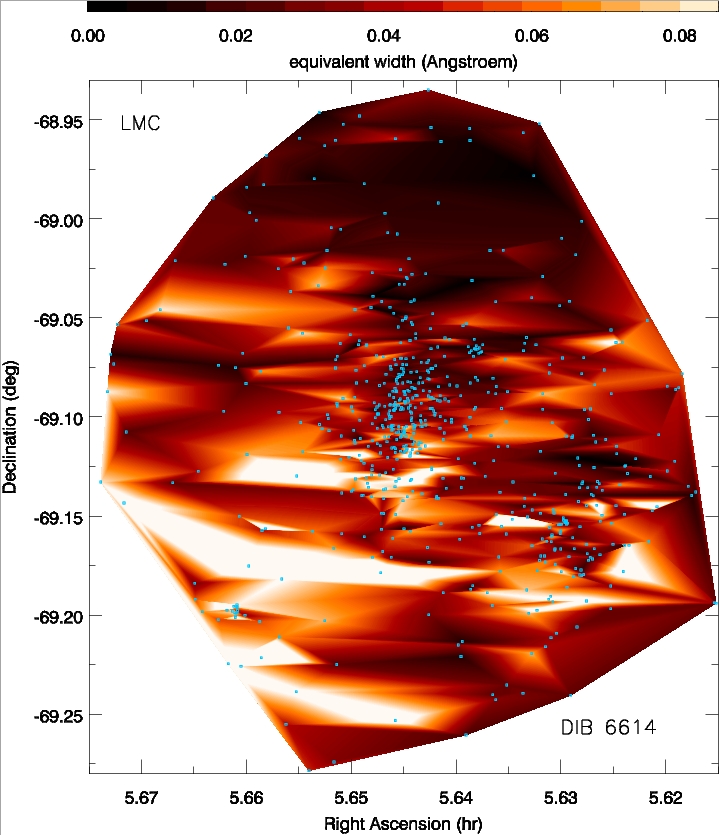,width=61mm}
\psfig{figure=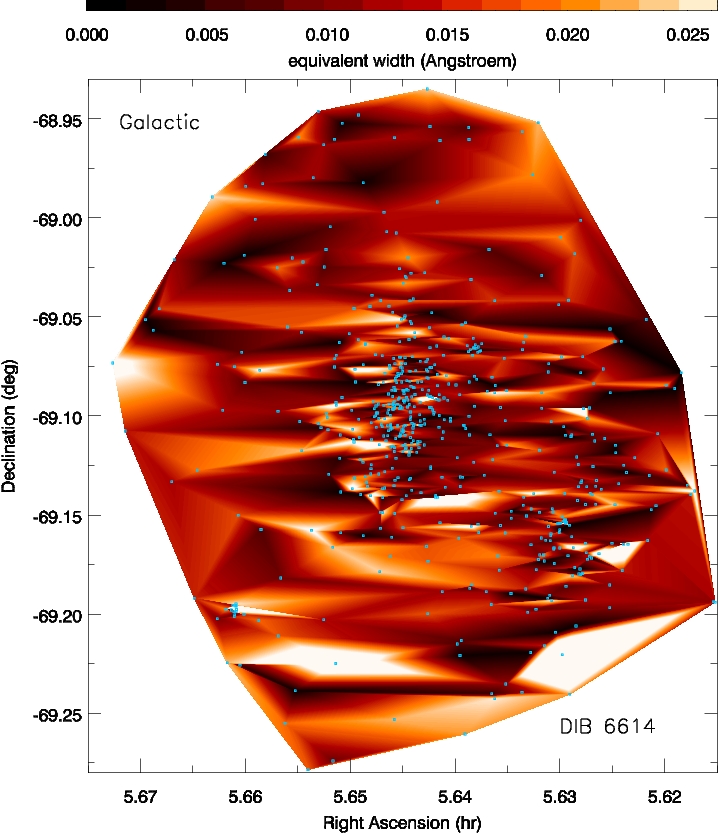,width=61mm}
\psfig{figure=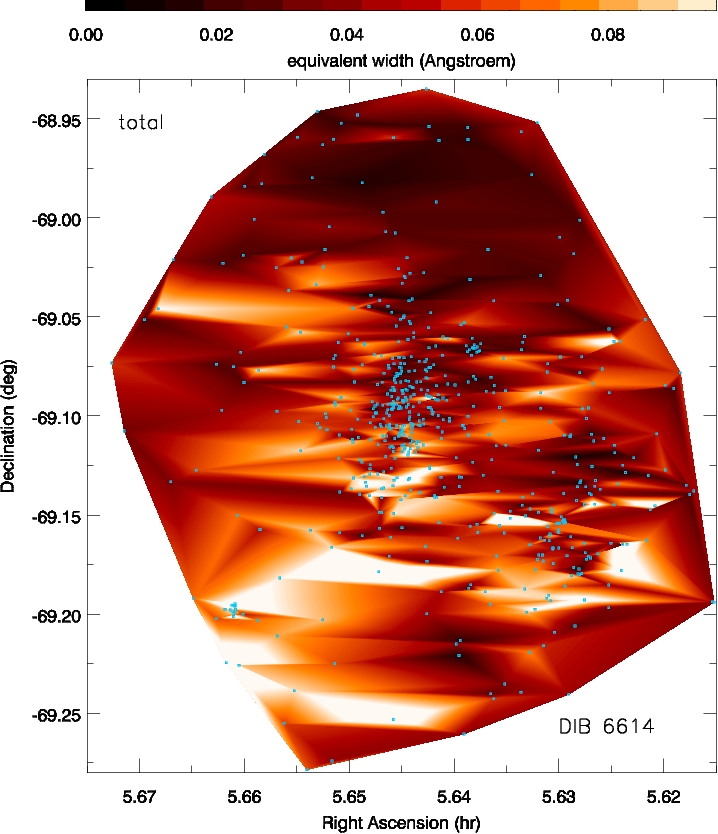,width=61mm}
}\vspace{2mm}\hbox{
\psfig{figure=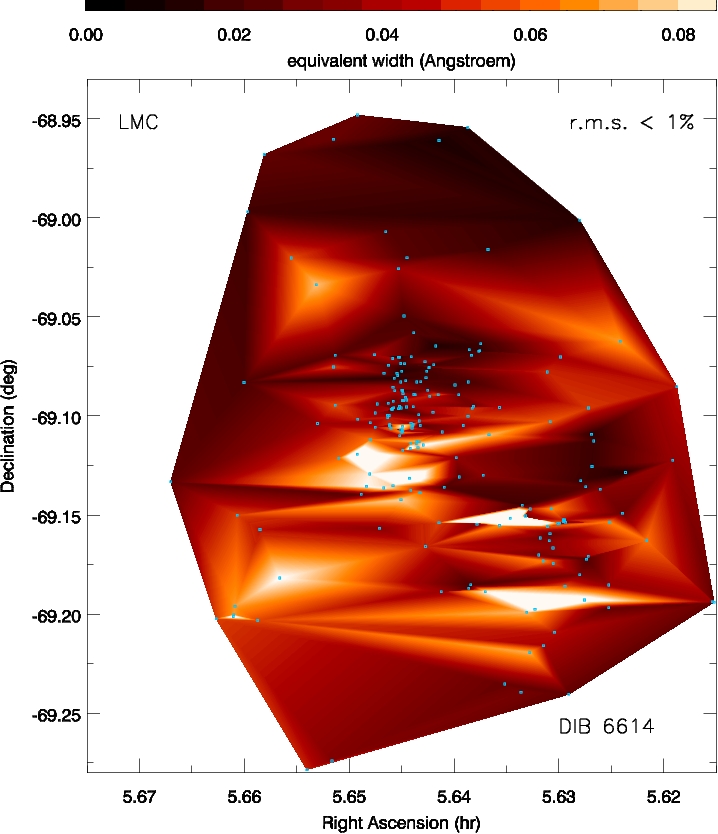,width=61mm}
\psfig{figure=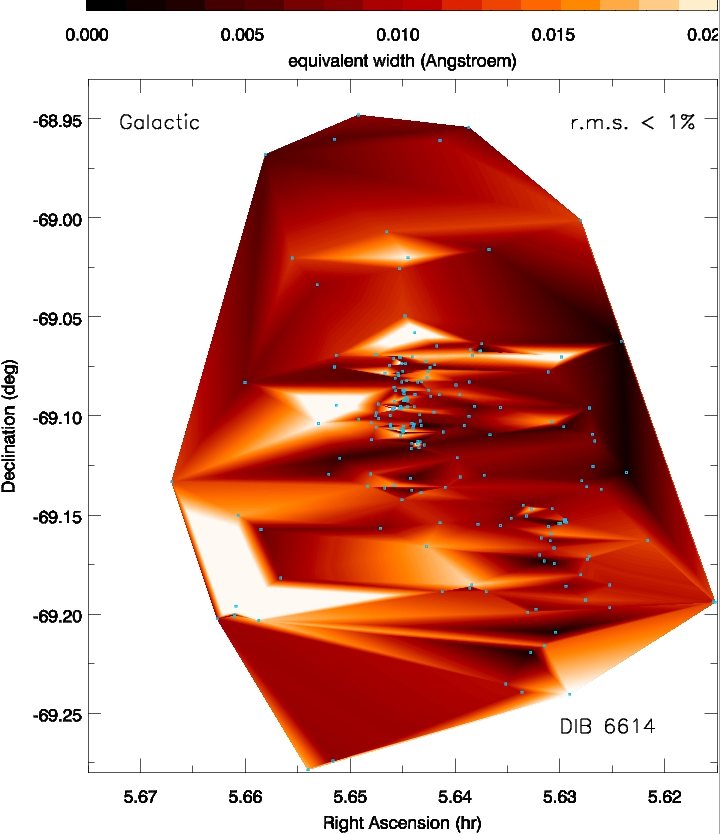,width=61mm}
\psfig{figure=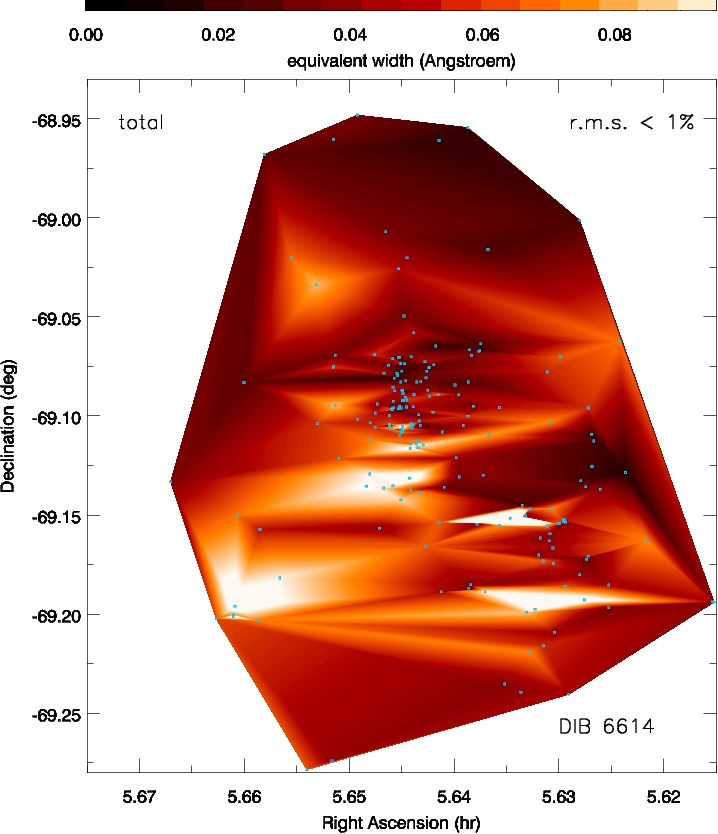,width=61mm}
}}}
\caption[]{Equivalent width maps of the 6614 \AA\ DIB absorption in the ({\it
Left:}) LMC component, ({\it Middle:}) Galactic component and ({\it Right:})
LMC + Galactic components, for ({\it Top:}) all and ({\it Bottom:}) the best
Medusa spectra where reasonable fits could be obtained. The maps cover an area
of $19^\prime({\rm RA})\times21^\prime({\rm Dec})$, which corresponds to
$0.28\times0.31$ kpc$^2$ at the distance of the LMC. The sightlines are marked
with little blue dots. The white box delineates an area which is displayed in
more detail in Fig.\ 7.}
\label{map3}
\end{figure*}

%
% FIGURE 7
%
\begin{figure*}
\centerline{\vbox{\hbox{
\psfig{figure=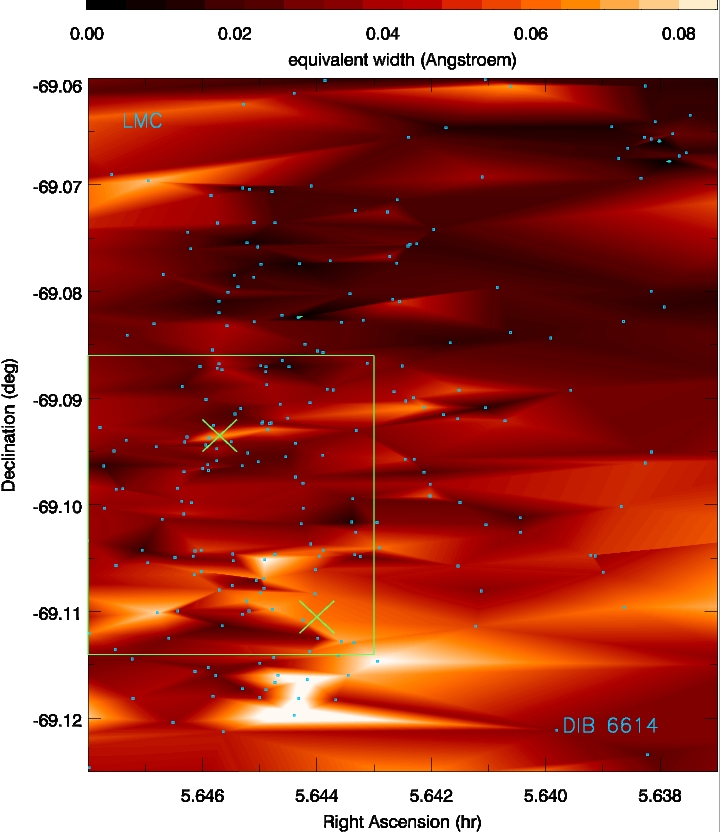,width=61mm}
\psfig{figure=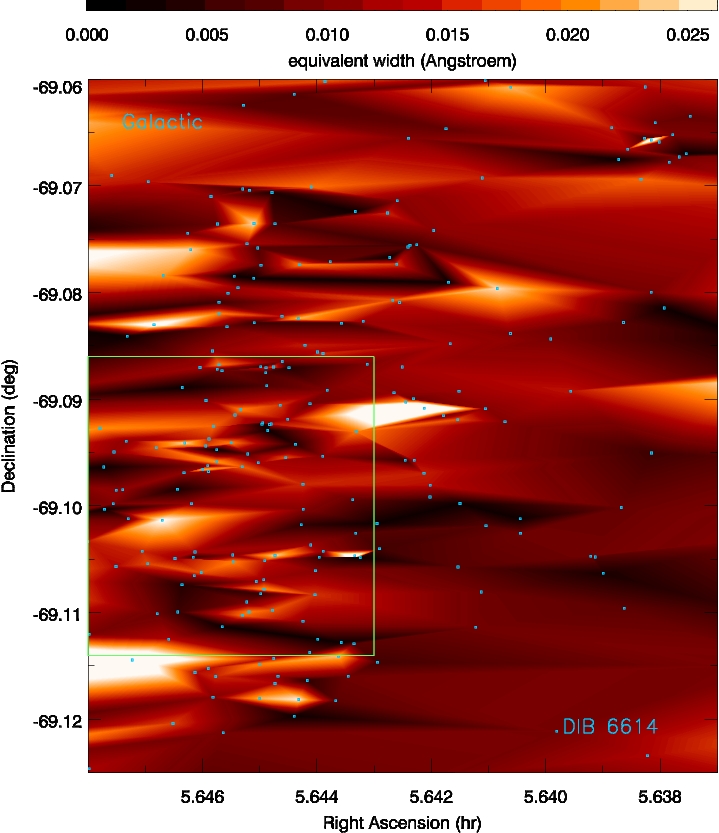,width=61mm}
\psfig{figure=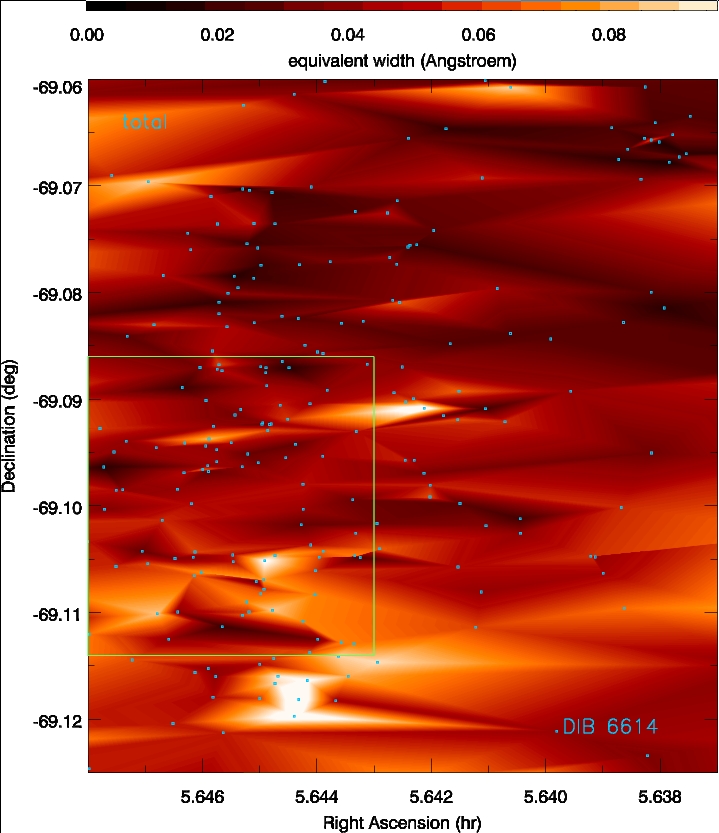,width=61mm}
}\vspace{2mm}\hbox{
\psfig{figure=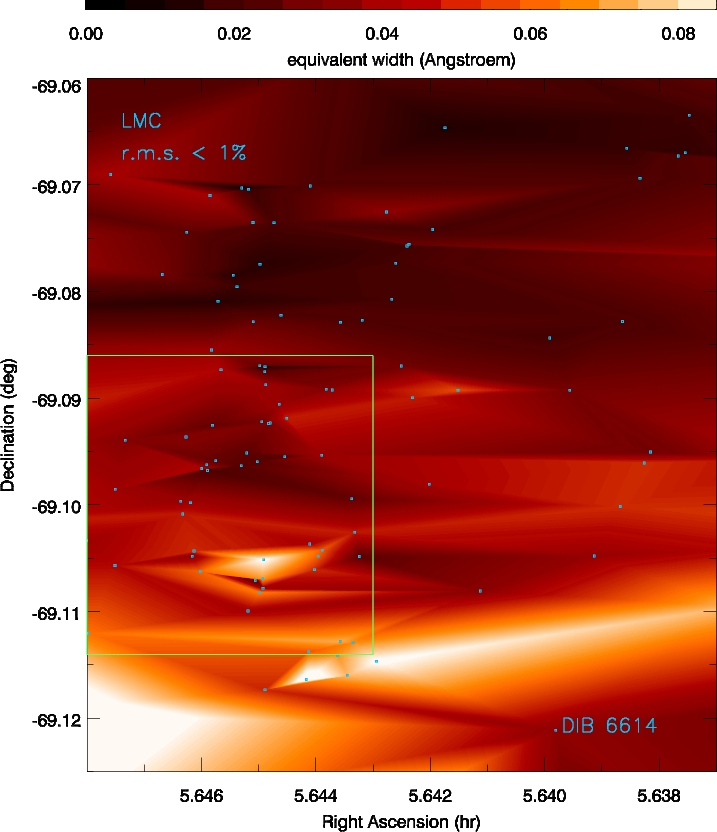,width=61mm}
\psfig{figure=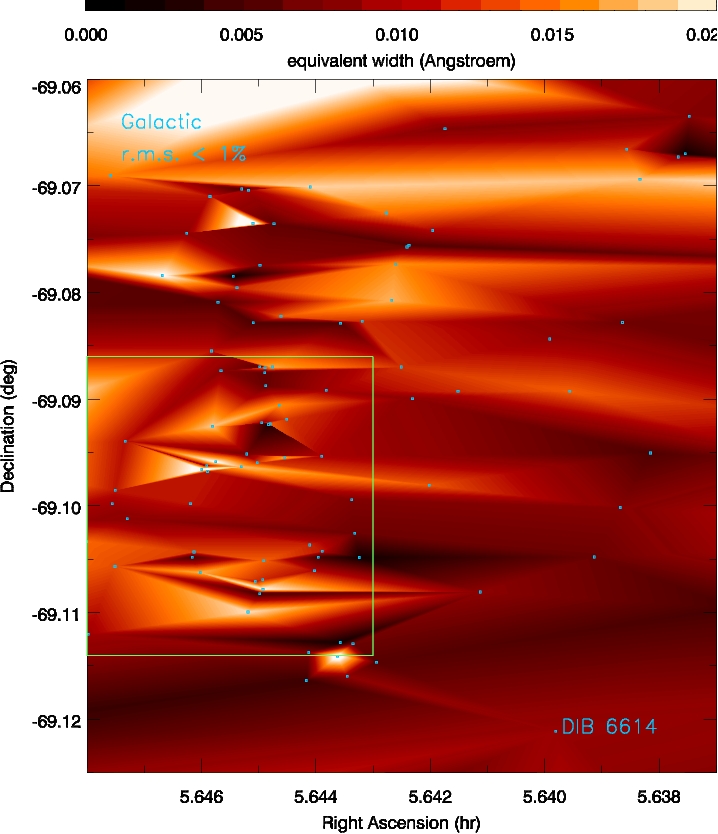,width=61mm}
\psfig{figure=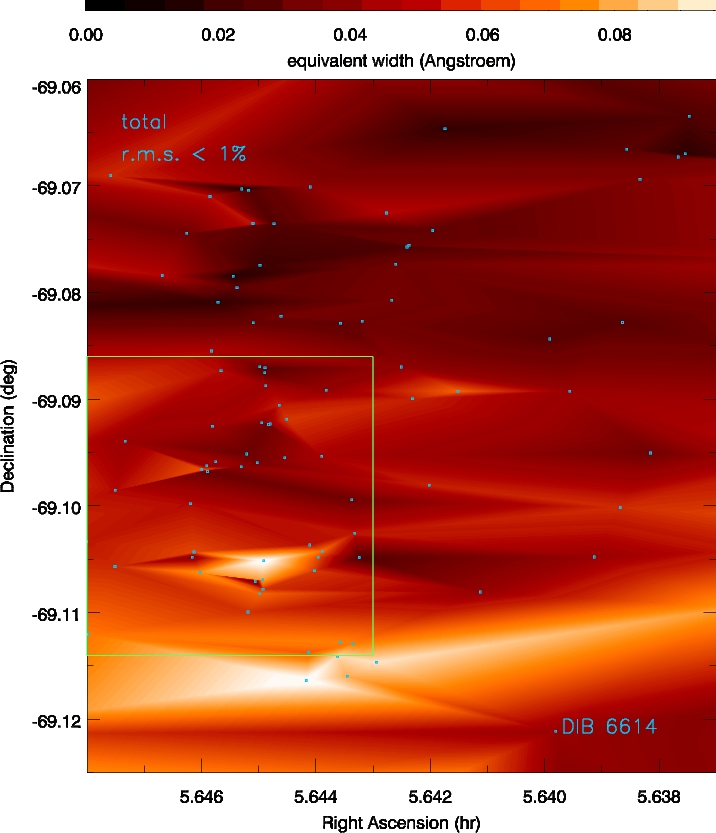,width=61mm}
}}}
\caption[]{As Fig.\ 6 but zoomed in on the densely sampled central region
measuring $3\rlap{.}^\prime5({\rm RA})\times3\rlap{.}^\prime9({\rm Dec})$,
which corresponds to $51\times57$ pc$^2$ at the distance of the LMC. The green
box delineates the area of the Na maps (Figs.\ 12 \& 13). The green crosses
mark two LMC clouds also seen in Na absorption (Fig.\ 12).}
\label{map4}
\end{figure*}

%
% FIGURE 8
%
\begin{figure*}
\centerline{\vbox{\hbox{
\psfig{figure=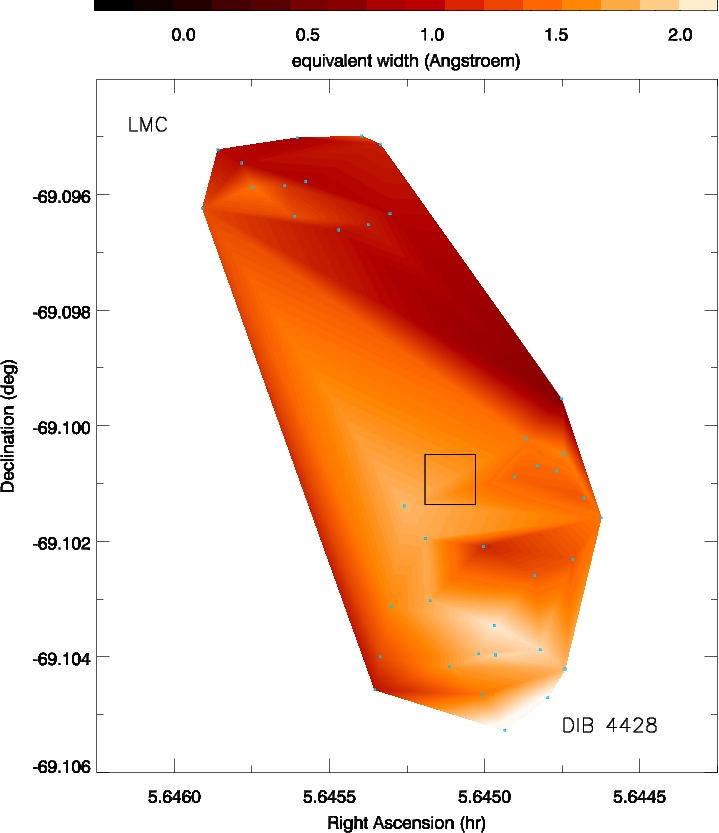,width=61mm}
\psfig{figure=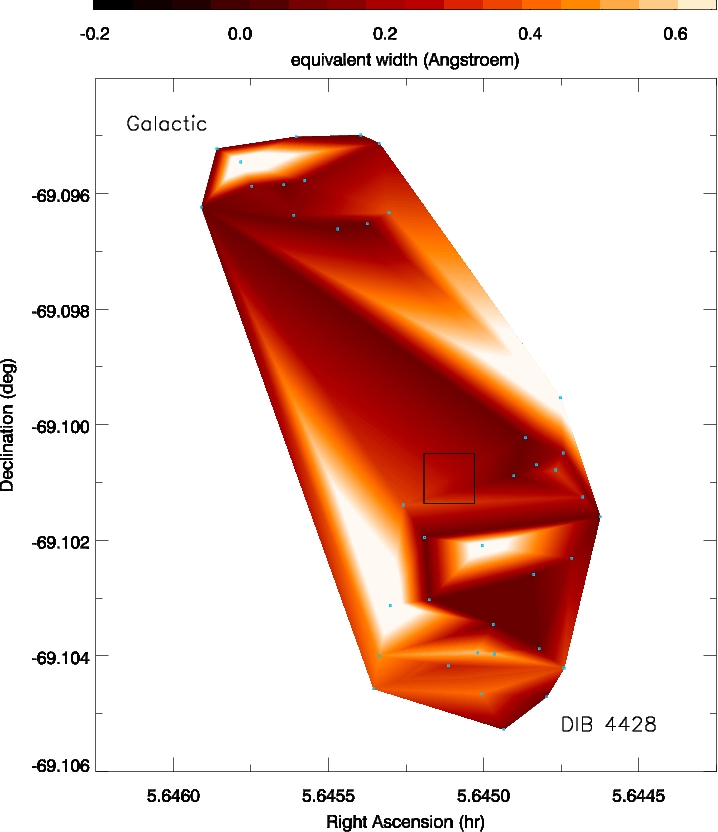,width=61mm}
\psfig{figure=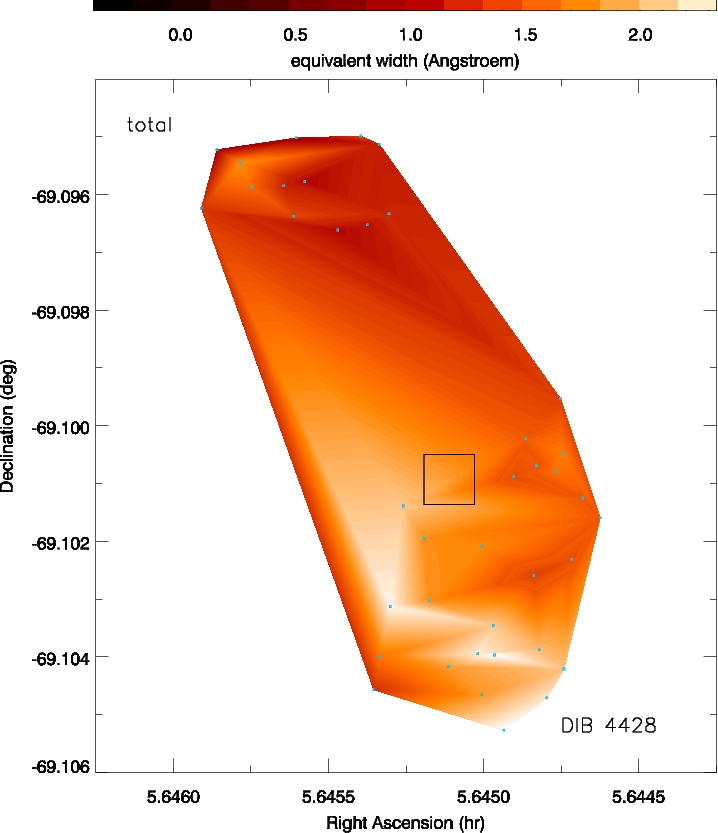,width=61mm}
}\vspace{2mm}\hbox{
\psfig{figure=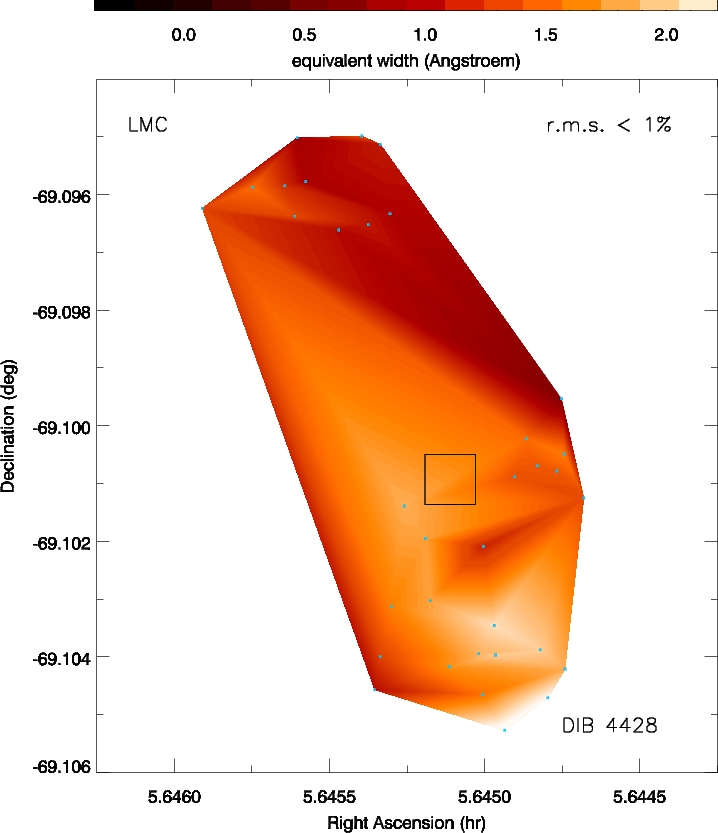,width=61mm}
\psfig{figure=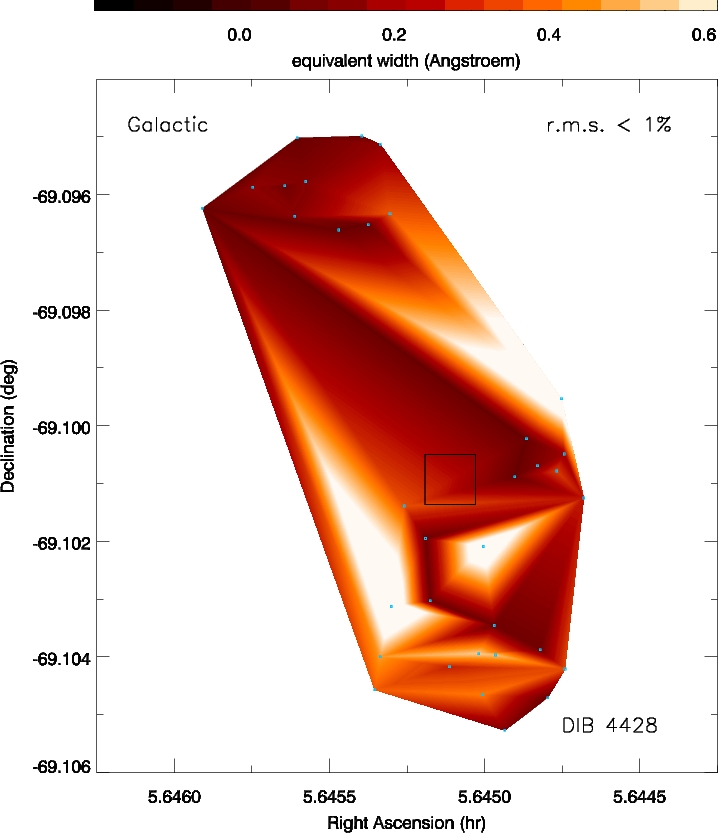,width=61mm}
\psfig{figure=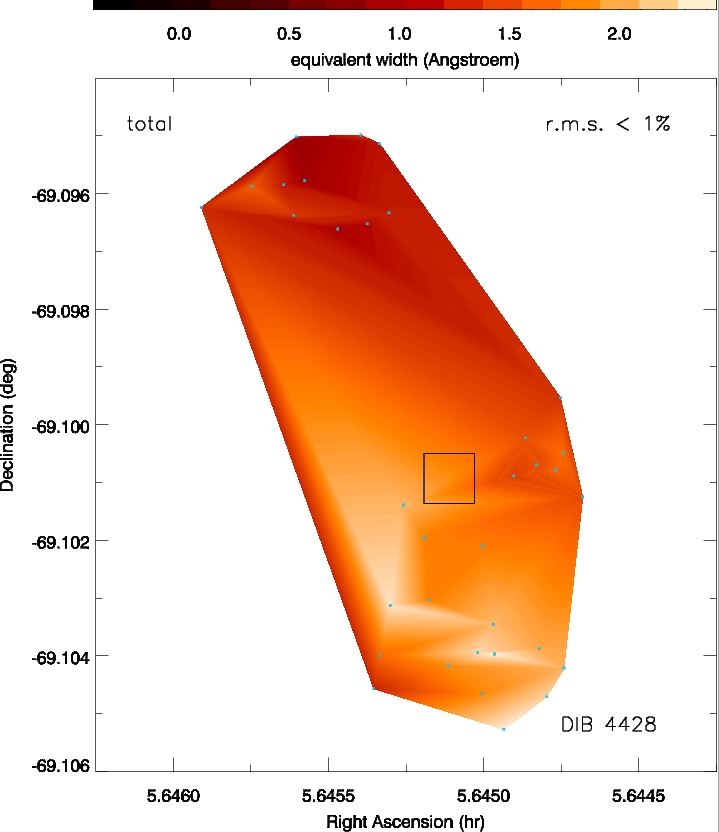,width=61mm}
}}}
\caption[]{As Fig.\ 4 but for the ARGUS spectra. The maps cover an area of
$38\rlap{.}^{\prime\prime}5({\rm RA})\times43^{\prime\prime}2({\rm Dec})$, which
corresponds to $9.3\times10.5$ pc$^2$ at the distance of the LMC. The sight
lines are marked with little blue dots. The large black square marks the core
of R\,136.}
\label{map5}
\end{figure*}

Maps of the 4428 \AA\ and 6614 \AA\ DIB equivalent width are presented in
Figs.\ 4 and 6 for the region covered by the Medusa spectra, in Figs.\ 5 and 7
zoomed in on the central region where the sampling is densest, and in Fig.\ 8
(ARGUS spectra, 4428 \AA\ only) for the region immediately surrounding the
R\,136 cluster as well as the cluster core itself. The maps are presented for
the LMC and Galactic components separately as well as jointly; we remind the
reader that the 6614 \AA\ DIB is separated completely but the 4428 \AA\ DIB is
a blend.

The two challenges inherent to presenting such maps are (i) the sparse and
non-uniform sampling and (ii) the difficulty of distinguishing small-scale
structure from noise fluctuations. The latter can be appreciated by presenting
maps using all reasonable fits to the spectral features and those making use
exclusively of the spectra with r.m.s.\ $<1$\%. Because the latter are based
on fewer sightlines the differences between the maps also give an impression
of the way the sparsity and non-uniformity of the sampling affect the maps.
The maps with only r.m.s.\ $<1$\% spectra miss some of the structures present
in the complete maps, but the reality of those structures is less certain;
there is not much one can do about that. On the other hand, because of the
sparse sampling and the ``filling up'' of the maps, structures in the r.m.s.\
$<1$\% maps may seem exacerbated in their extent. On balance, the 4428 \AA\
and 6614 \AA\ DIB maps of the LMC component show more similarity when
accepting all reasonable fits than when limiting oneself to the r.m.s.\ $<1$\%
fits, because of the lack of commonality in sightlines. Intriguingly, this in
itself would suggest that real structure is presented at scales below the
sampling of the ``all-inclusive'' maps, i.e.\ at several pc.

The LMC and Galactic maps look totally different. This is to be expected, as
there is no physical connection between these two disparate regions in the
universe. However, the discerning eye may notice a certain level of
anti-correlation between the strongest features in the Galactic maps of the
4428 \AA\ DIB and rather weak features at the corresponding positions in the
LMC maps -- a consequence of the difficulty of disentangling the two
components in this broad DIB. This does not necessarily mean that the Galactic
feature is unreliable, though its strength must be taken with caution. A
corresponding structure in the 6614 \AA\ DIB would lend credit to its reality.

%------------------------------------------------------------------------- 3.2
\subsection{Correlation between DIBs}

%
% FIGURE 9
%
\begin{figure*}
\centerline{\hbox{
\psfig{figure=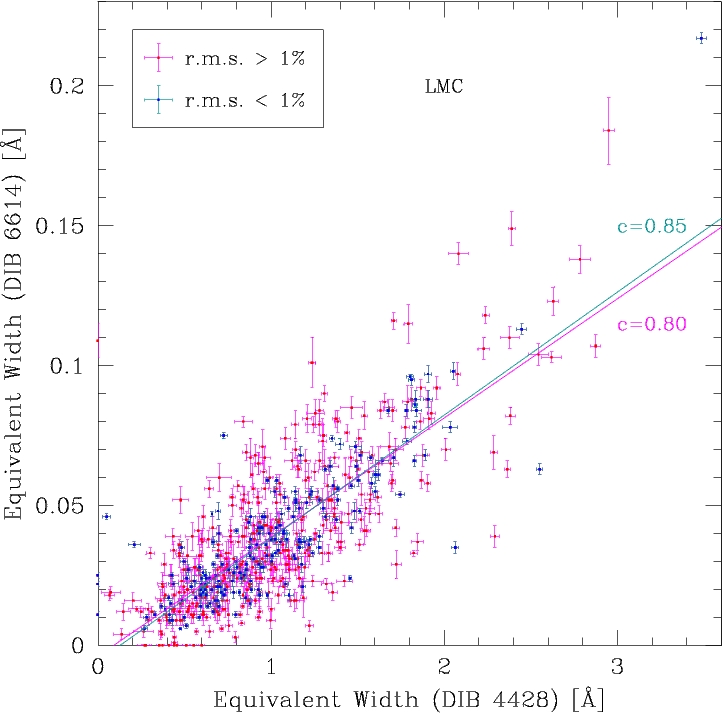,width=61mm}
\psfig{figure=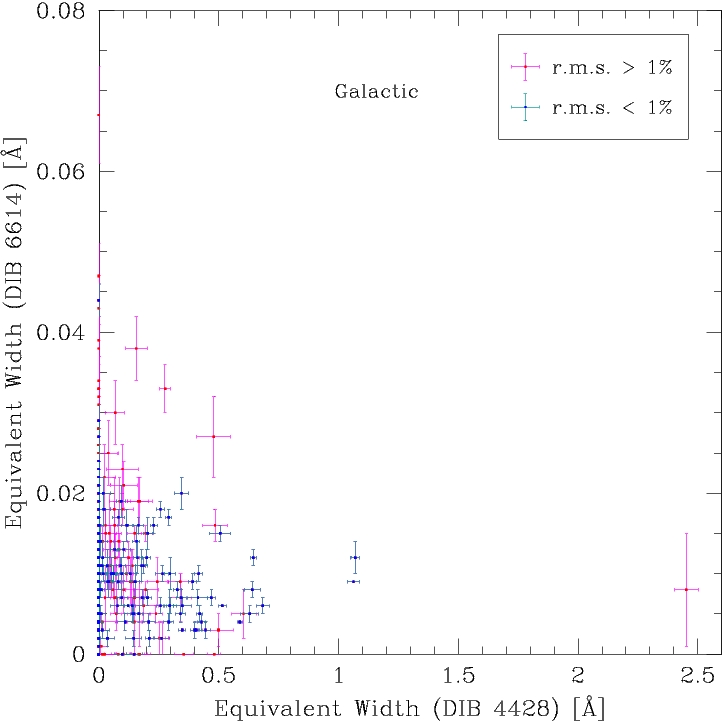,width=61mm}
\psfig{figure=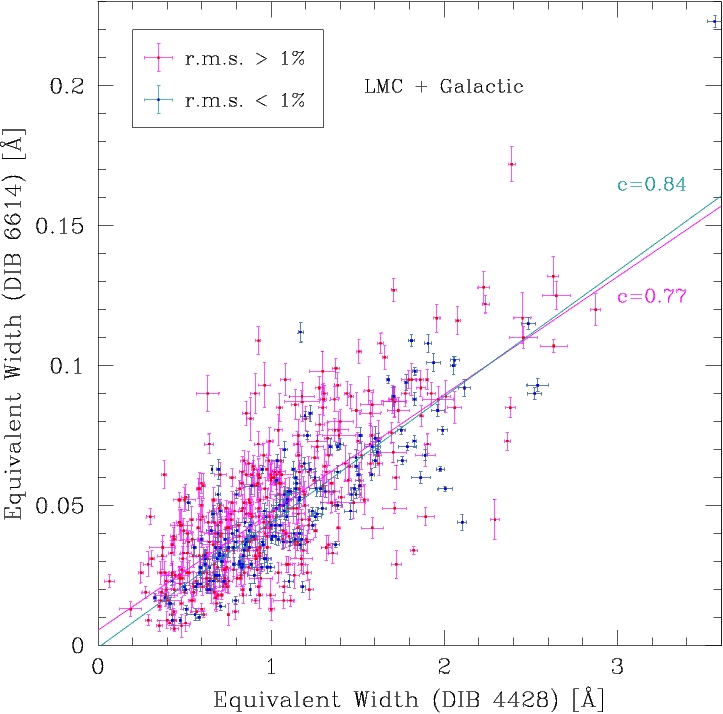,width=61mm}
}}
\caption[]{Correlation between the 4428 \AA\ and 6614 \AA\ DIBs, for the ({\it
Left:}) LMC component, ({\it Middle:}) Galactic component and ({\it Right:})
LMC + Galactic components. A distinction is made between acceptable fits and
the best fits (with r.m.s.\ $<1$\%). Linear regression lines are overplotted,
and annotated with the linear correlation coefficient.}
\label{correlation1}
\end{figure*}

As noted in the maps, the equivalent widths of the 4428 \AA\ and 6614 \AA\
DIBs are correlated. This is very clear for the LMC component (Fig.\ 9, left,
correlation coefficient $c\sim0.8$) though not so for the Galactic component
(Fig.\ 9, middle, correlation coefficient $c\sim-0.1$).

%
% FIGURE 10
%
\begin{figure*}
\centerline{\hbox{
\psfig{figure=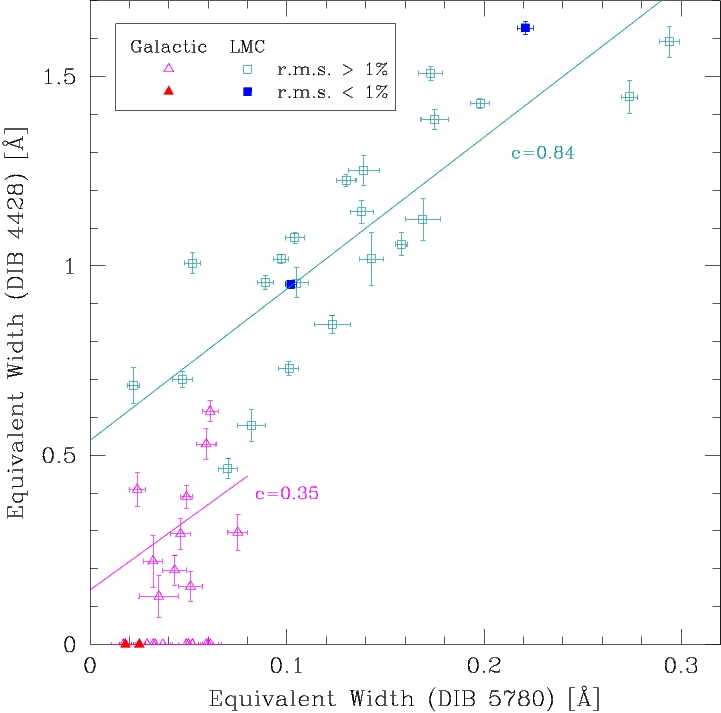,width=61mm}
\psfig{figure=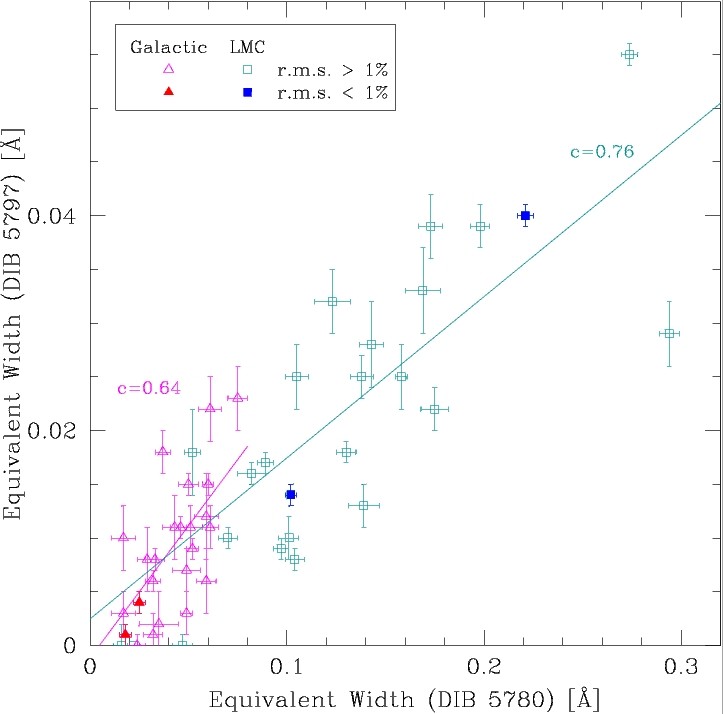,width=61mm}
\psfig{figure=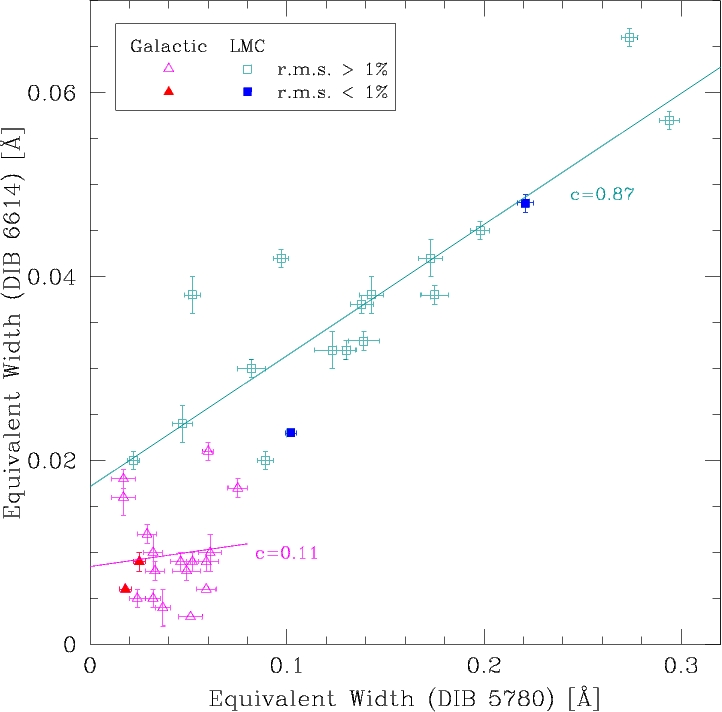,width=61mm}
}}
\caption[]{Correlation between the ({\it Left:}) 4428 \AA\ DIB, ({\it
Middle:}) 5797 \AA\ DIB and ({\it Right:}) 6614 \AA\ DIB on the one hand, and
the 5780 \AA\ DIB on the other. The Galactic component is plotted with open
magenta triangles and the LMC component with open cyan squares; the best fits
(with r.m.s.\ $<1$\%) for both DIBs are highlighted in solid red triangles and
solid blue squares for the Galactic and LMC components, respectively. Linear
regression lines are overplotted, and annotated with the linear correlation
coefficient.}
\label{correlation2}
\end{figure*}

The equivalent width of the 4428 \AA\ DIB is also correlated with that of the
5780 \AA\ DIB (Fig.\ 10, left), but not perfectly. In the LMC, the 4428 \AA\
DIB is already obvious when the 5780 \AA\ DIB is still weak, implying the
carrier of the former is more abundant and suggesting its carrier may already
-- or still -- be present when the carrier of the latter is not. One can also
notice that -- at least in the weak 5780 \AA\ DIB regime -- the 4428 \AA\ DIB
is a lot stronger in the LMC than in the Milky Way, which means that either
the carrier of the 4428 \AA\ DIB is relatively abundant in the LMC or the
carrier of the 5780 \AA\ DIB is depleted in the LMC (or both). These
conclusions might be affected by the difficulty in separating the LMC and
Galactic components in the broad 4428 \AA\ DIB. However, they are corroborated
by the behaviour of the 6614 \AA\ DIB where no such blending occurs (Fig.\ 10,
right); the inferences with regard to the carrier of the 4428 \AA\ DIB may
therefore also hold for the carrier of the 6614 \AA\ DIB. The equivalent width
of the 5797 \AA\ DIB is rather well correlated with that of the 5780 \AA\ DIB
(Fig.\ 10, middle).

The growth of the equivalent width of the 4428 \AA\ DIB appears to slow down
as the equivalent width of the 5780 \AA\ DIB grows. This might be understood
if the ISM becomes depleted of the building blocks that make up the carrier of
the 4428 \AA\ DIB, or if the carrier of the 4428 \AA\ DIB itself is being
depleted in regions where the 5780 \AA\ DIB is particularly strong.

%------------------------------------------------------------------------- 3.3
\subsection{Correlation between DIBs and atomic gas}

%
% FIGURE 11
%
\begin{figure*}
\centerline{\hbox{
\psfig{figure=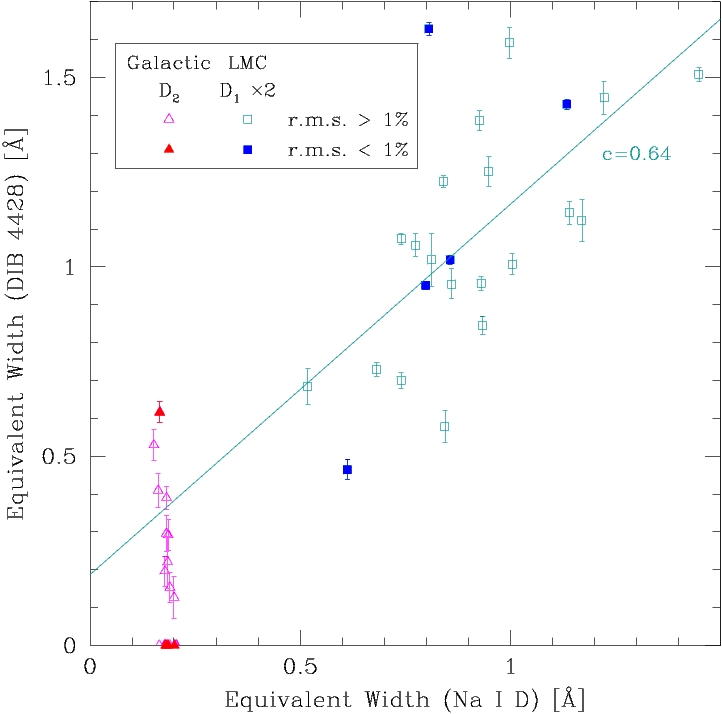,width=61mm}
\psfig{figure=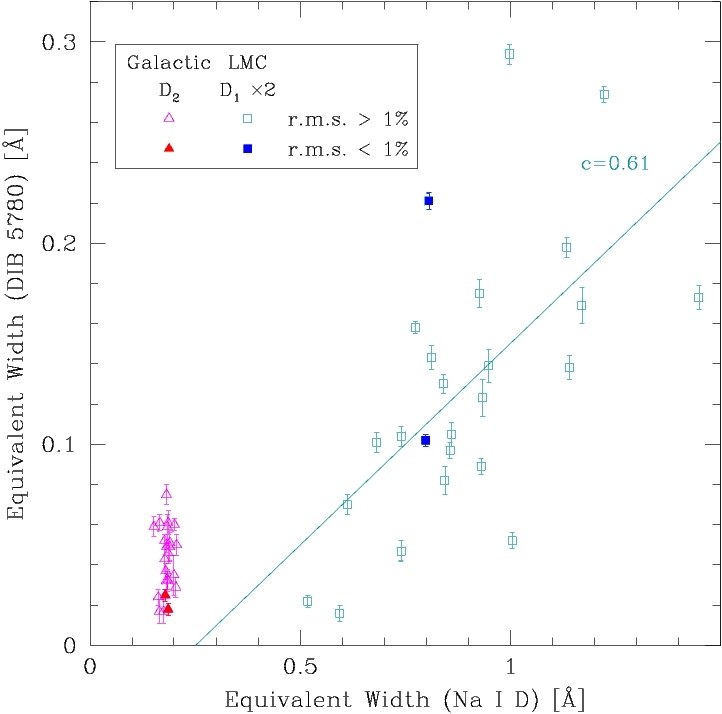,width=61mm}
\psfig{figure=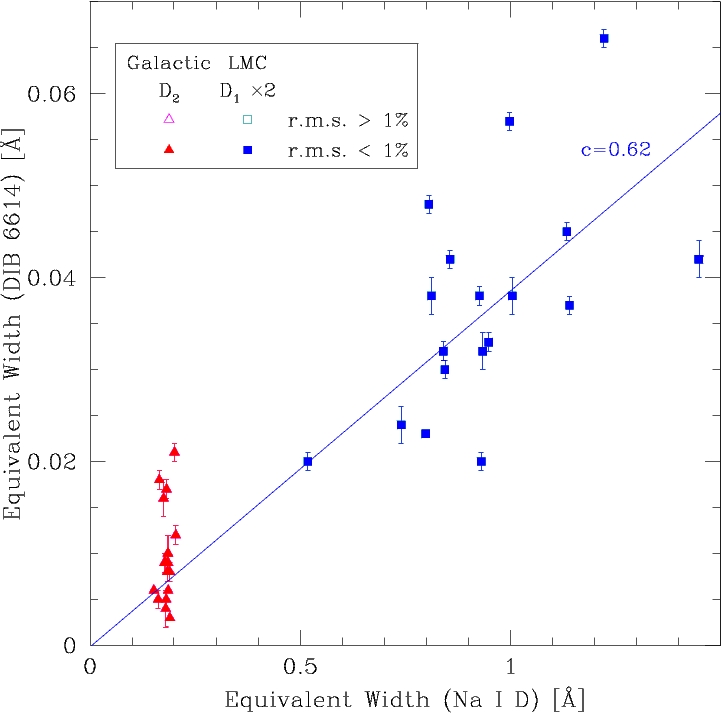,width=61mm}
}}
\caption[]{Correlation between the ({\it Left:}) 4428 \AA\ DIB, ({\it
Middle:}) 5780 \AA\ DIB and ({\it Right:}) 6614 \AA\ DIB on the one hand, and
the Na\,{\sc i}\,D line on the other. The Galactic component (D$_2$
transition) is plotted with open magenta triangles and the LMC component
(D$_1$ transition, multiplied by the ratio of intrinsic strengths of the D$_2$
and D$_1$ transition) with open cyan squares; the best fits (with r.m.s.\
$<1$\%) for both DIBs are highlighted in solid red triangles and solid blue
squares for the Galactic and LMC components, respectively. Linear regression
lines are overplotted, and annotated with the linear correlation coefficient.}
\label{correlation3}
\end{figure*}

The LMC component of the DIBs shows a clear correlation with the strength of
Na\,{\sc i}\,D absorption (Fig.\ 11) despite the latter suffering from
saturation effects. If the equivalent width of the Na\,{\sc i}\,D$_1$ line in
the LMC component largely reflects the number of kinematically separated
clouds then this could mean that the DIBs are associated with each of these
clouds as opposed to certain clouds but not others. However, at the same Na
equivalent width the DIBs can display variations in excess of a factor two
(see for example the two very good spectra of the 5780 \AA\ DIB in Fig.\ 11).

The Galactic component of the Na\,{\sc i}\,D absorption does not exhibit much
variation despite significant scatter in the DIB absorption; this might be due
to saturation of the Na\,{\sc i}\,D$_2$ line. If anything there seems to be
some indication of an {\it anti}-correlation between the 4428 \AA\ DIB and
Na\,{\sc i}\,D, but we discard this as this is the most difficult DIB for
which to separate the weak Galactic component from the stronger LMC component
and the other DIBs do not show such anti-correlation.

There appears to be a threshold in Na column density below which the 5780 \AA\
DIB is not detected in the LMC; while we have no sightlines with such low Na
column densities it appears evident from extrapolation (Fig.\ 11, middle).
This is certainly not evident for the 4428 \AA\ DIB (Fig.\ 11, left); the
behaviour of the 6614 \AA\ DIB in this respect is in between that of the 4428
and 5780 \AA\ DIBs. This suggests that, as the neutral atomic gas abundance
grows, first the carrier of the 4428 \AA\ DIB appears subsequently followed by
that of the 6614 \AA\ DIB and finally that of the 5780 \AA\ DIB. This can be
reconciled with the scenario in which the carrier of the 5780 \AA\ DIB resides
in the ``skin'' of clouds as such clouds are not prevalent in relatively warm
and diffuse gas. The carrier of the 4428 \AA\ DIB may then, by inference,
reside in the inter-cloud medium (as well).

As noted before, the Na\,{\sc i}\,D absorption shows a great deal of
complexity both in kinematic and spatial respect. To visualize this, we
display ``channel maps'' of $\sim5$ km s$^{-1}$ wide slices through the D$_1$
line profile of the LMC component (Fig.\ 12). One may note that the region in
the immediate vicinity of R\,136 is devoid of gas compared to its surroundings
at all velocities except for the high-speed gas cloud at $\sim235$ km s$^{-1}$
which is centered exactly at it. This suggests that the stellar winds
emanating from R\,136 have carved a cavity of roughly a dozen pc across, and
spat out the blue-shifted gas which is now moving outward at a speed of nearly
40 km s$^{-1}$.

Not surprisingly, it is the strongest Na\,{\sc i}\,D$_1$ structures that are
easiest to associate with features in the DIB maps (Figs.\ 5 \& 7): the
discrete cloud around 276 km s$^{-1}$ at (RA,Dec)
$\simeq(5\rlap{.}^\circ6457,-69\rlap{.}^\circ094)$ can be seen as an equally
discrete structure in both the 4428 \AA\ and 6614 \AA\ DIB absorption; the
absorption in the south--west corner of the Na\,{\sc i}\,D$_1$ map around 276
km s$^{-1}$ is associated with strong 4428 \AA\ DIB absorption around (RA,Dec)
$\simeq(5\rlap{.}^\circ644,-69\rlap{.}^\circ11)$; and the filament around 255
km s$^{-1}$ stretching towards the West from (RA,Dec)
$\simeq(5\rlap{.}^\circ647,-69\rlap{.}^\circ107)$ has a counterpart in the 4428
\AA\ DIB absorption as well.

The Galactic component is split into two slices: one centered at 9 km s$^{-1}$
and 7 km s$^{-1}$ wide, and another centered at 24 km s$^{-1}$ and 15 km
s$^{-1}$ wide (Fig.\ 13). Structure is seen in both kinematic components on
scales of $<1^\prime$, or 0.01 pc at 40 pc distance. We discuss this further
in Section 4.2.2. There is no clear association between individual features in
the Na\,{\sc i}\,D maps and the DIBs (Figs.\ 4--8), so we cannot conclude
whether the carriers of the DIBs are more prevalent in the nearby or more
distant gas.

%
% FIGURE 12
%
\begin{figure*}
\centerline{\psfig{figure=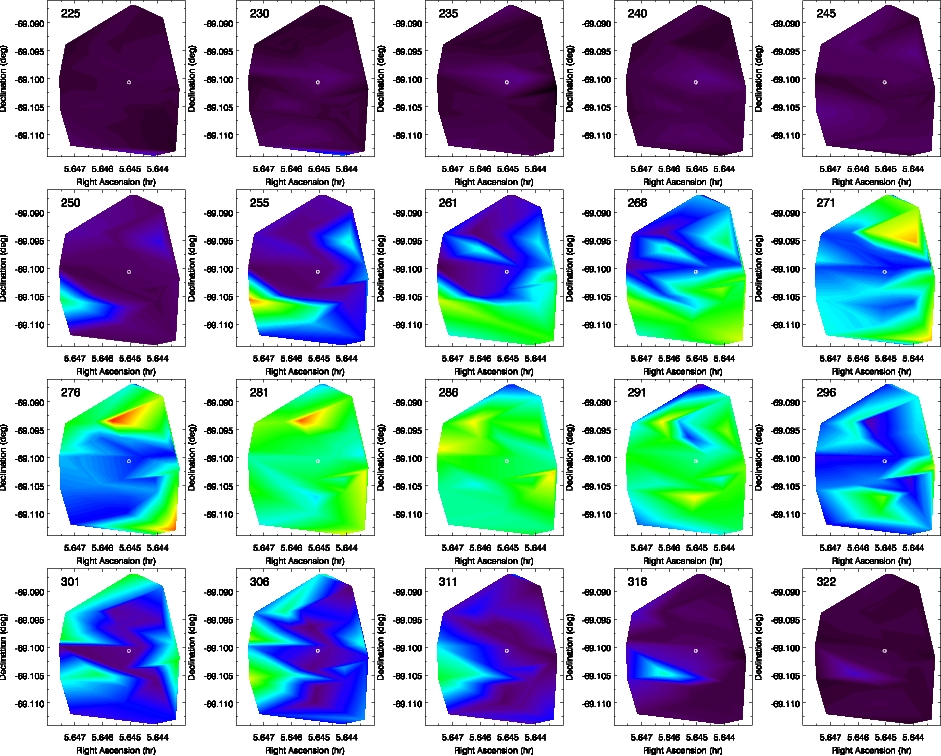,width=183mm}}
\caption[]{Velocity slices through the LMC component of the Na\,{\sc i}\,D$_1$
absorption in the UVES spectra, using a ``rainbow'' intensity scale (from
black where there is no absorption, through purple/blue, then green/yellow, to
red at the maximum detected absorption level). The centre velocity of each
$\sim5$ km s$^{-1}$ wide slice is printed in each panel. The centre of R\,136
is marked with a small white circle.}
\label{sodium1}
\end{figure*}

%------------------------------------------------------------------------- 3.4
\subsection{Correlation between DIBs and visual extinction}

Because no distinction can be made between the LMC and Galactic components
contributing to the interstellar attenuation due to dust grains, we only
inspect the correlation between the {\it total} (i.e.\ LMC + Galactic) DIB
absorption and visual extinction, A$_{\rm V}$. The latter was derived from a
comparison between photometry and spectral modelling of a subset of the VFTS
sample -- mostly O-type stars (Ma\'{\i}z Apell\'aniz et al.\ in prep.). There
is quite a good correlation (Fig.\ 14), at least for $A_{\rm V}<2$ mag. The
correlation seems better for the 4428 \AA\ DIB than for the 6614 \AA\ DIB.
Strong DIBs are only seen through dusty sightlines, but the reverse is not
true: some dusty sightlines have clear but not particularly strong DIBs.

The 4428, 5780 and 5797 \AA\ DIBs as well as Na\,{\sc i}\,D absorption are all
correlated with A$_{\rm V}$ (Fig.\ 15, which only shows sightlines in common,
i.e.\ in the direction of R\,136). However, when extrapolating towards the
origin, the Na absorption does not vanish when A$_{\rm V}$ does, whilst the
5780 and 5797 \AA\ DIBs seem to require a threshold in A$_{\rm V}$ of about 0.6
mag; the behaviour of the 4428 \AA\ DIB is consistent with it vanishing when
A$_{\rm V}$ does, especially given that the full sample (Fig.\ 14) does not
support an extinction threshold for this ubiquitous DIB.

%=========================================================================== 4
\section{Discussion}

%------------------------------------------------------------------------- 4.1
\subsection{The nature of the carriers of DIBs}

While we use the DIBs to map the ISM this is done without knowing the nature
of the carriers of these absorption bands, but by doing so we gain insight
into their behaviour which may help constrain their possible character. Clues
may be had from the shapes of the bands, from correlations between the bands
and between bands and other ISM tracers such as atomic absorption lines or the
attenuation or emission by dust grains, or from the spatial distribution of
the DIB absorption strength in relation to physical structures in the ISM.

To start with the latter approach, we compare the (Medusa, total) 4428 \AA\
DIB map with an optical and a far-IR view of the same region (Fig.\ 16). The
optical picture is a composite of broad-band images showing mainly stars and
some reflection nebulosity, and narrow-band images centered on the [O\,{\sc
iii}] (green) and H$\alpha$ (red) emission lines. Besides stars -- in
particular the four clusters R\,136, NGC\,2060, Hodge\,301 and SL\,639 -- the
optical appearance features emitting or reflecting (yellow/brown) nebulosity
as well as dark patches and filaments obscuring optical light from stars and
nebulosity behind them. The far-IR image is a broad-band 70-$\mu$m image
obtained with the {\it Spitzer} Space Telescope for the SAGE-LMC legacy survey
(Meixner et al.\ 2006); it shows thermal emission from cool dust. Note that
the DIB map is solely based on material absorbing light from the stars behind
them: clouds of DIB carriers lying behind stars will remain unnoticed. At the
same time, our sample is biased against probing very dense clouds as the high
visual extinction would render the stars too faint. That said, there is
remarkably little correspondence between the DIB map and the dust map. The
already imperfect link between DIB absorption and visual extinction is lost
when the effects of geometry (spatial distribution of stars and dust) and
variations in dust temperature and illumination (hence IR surface brightness)
are folded in.

More striking is the lack of DIBs at the exact positions of the four star
clusters, but their prominence adjacent to (or even surrounding) at least
three of them. While there is no clear correspondence between the DIB map and
the diffuse optical emission, the lack of DIBs in the north--west corresponds
to regions where the [O\,{\sc iii}] emission is strong; strong DIBs are found
instead in the H$\alpha$-dominated regions in the direction of the molecular
ridge to the south of the Tarantula Nebula (cf.\ Wong et al.\ 2011). The
north--west is littered with superbubbles created by multiple supernovae
predating the formation of R\,136 (Kim et al.\ 1999); the intense [O\,{\sc
iii}] emission might arise from shocks at the rims of those bubbles, and the
bubbles themselves might be largely ``empty''. This suggests that the carriers
of the DIBs -- and/or their excitation -- reside in relatively diffuse, weakly
ionized gas and that they are diminished in dense, cold clouds as well as in
hot bubbles and when directly irradiated by hot stars. This setting the stage,
we now look in more detail into the properties and behaviour of the DIBs.

%....................................................................... 4.1.1
\subsubsection{Band profiles}

The profile of the 4428 \AA\ DIB in our data was better represented with a
Lorentzian profile than with a Gaussian profile, in accordance with the
findings by Snow et al.\ (2002b) who interpreted this as evidence for a
molecular carrier which would give rise to the damping wings of the profile.
The 6614 \AA\ DIB in our data was better represented with a Gaussian profile,
though one may discern in the combined spectrum (Fig.\ 2) that the
short-wavelength edge is steeper and more abrupt than the long-wavelength
edge, and that the core seems to have a ``shoulder''. These observations are
in agreement with the substructure noted by Sarre et al.\ (1995) and Cami et
al.\ (2004) in high-resolution spectra. In this regard, the 4428 and 6614 \AA\
DIBs display similar characteristics both in the LMC and in the Milky Way.

The FWHM of the 4428 \AA\ DIB was fixed to 20 \AA; we saw no indications that
individual profiles of strong absorption deviated from this at any significant
level. Such width is typical of Galactic sightlines as probed by, e.g., Snow
et al.\ (2002b) and Hobbs et al.\ (2008, 2009) who found slightly (of order
10\%) smaller and larger values, respectively. This suggests a great deal of
uniformity of the carrier and insensitivity of the excitation of the band to
environmental conditions, as also noted by Snow et al. The FWHMs of the 5780,
5797 and 6614 \AA\ DIBs were measured for both the LMC and Galactic sightlines
in our sample, and the results are listed in Table 3. For the 6614 \AA\ DIB we
had the luxury of numbers to limit the tabulated values to those measured in
the best spectra with r.m.s.\ $<1$\%; the median FWHM for spectra with r.m.s.\
$>1$\% is $\sim10$\% smaller but with a somewhat larger standard deviation (in
all cases we omitted values of zero). As an independent reference, the values
in Hobbs et al.\ (2009) based on one spectrum are 2.14, 0.91 and 1.08 \AA\ for
the 5780, 5797 and 6614 \AA\ DIBs, i.e.\ similar to both components in our
sample. Nonetheless, one could argue that, if anything, the profiles in the
LMC are narrower than those in the Milky Way -- especially considering that
the kinematic broadening may be more important in the LMC than in (at least
our) Galactic sightlines. Indeed, we have seen clear evidence of strong
absorption from sodium over more than 40 km s$^{-1}$, which in itself would
yield a DIB profile FWHM of 0.8--0.9 \AA. The latter cannot be true as the
FWHM of the 5797 \AA\ DIB does not exceed that value and the intrinsic width
of the DIBs is certainly more than a few tenths of \AA\ (see below for further
discussion). If impurities in the carrier cause additional sub-structure,
e.g., as seen in the 5797 \AA\ by Kerr et al.\ (1998) and/or broadening then
one could envisage a scenario in which such impurities are suppressed in a
metal-depleted environment such as that of the LMC as compared to that of the
Milky Way Disc (Milky Way Halo clouds being metal-poor).

%....................................................................... 4.1.2
\subsubsection{Correlations}

The 4428, 5780, 5797 and 6614 \AA\ DIBs are all correlated with one another.
Moutou et al.\ (1999), in spectra of nearby stars, found that the 6614 \AA\
DIB correlates well with the 5780 and 5797 \AA\ DIBs but not with the 4428
\AA\ DIB. In contrast, our data shows a rather good correlation between the
6614 and 4428 \AA\ DIBs (Fig.\ 9). There is some scatter in this relation, and
in particular there seem to be a number of LMC sightlines with relatively weak
6614 \AA\ DIB for the observed strength of the 4428 \AA\ DIB in that
direction. Looking at where these are distributed on the sky (Fig.\ 17) one
may notice that these sightlines are found near to the two main OB
associations, viz.\ R\,136 and NGC\,2060 (see Fig.\ 16), and the northern part
of the field, but not in the molecular cloud complexes to the south. This
could signify a DIB selection process driven by UV irradiation, with the 4428
\AA\ DIB being more resilient in the harsh radiation environment even though
it too is suppressed in those environments (Fig.\ 16).

The ratio between the equivalent widths of the 5797 and 5780 \AA\ DIBs,
$EW(5797)/EW(5780)=0.21$ in the Galactic sightlines of our own sample, and
0.15 in the LMC, i.e.\ firmly typical of $\sigma$-type clouds characterised by
a harsh radiation field. This is not surprising as the high Galactic latitude
sightlines probe hot gas in the Local Bubble similar to the extra-planar gas
of the Milky Way for which this was already demonstrated (van Loon et al.\
2009), and the Tarantula Nebula is exposed to the irradiation by many young
massive O-type stars. The data in Welty et al.\ (2006) suggest
$EW(5797)/EW(5780)=0.44$, 0.40 and 0.37 for Galactic, LMC and SMC sightlines,
respectively. This suggests generally stronger 5780 and/or weaker 5797 \AA\
DIBs for decreasing metallicity, but our lower values in both the Galactic and
LMC sightlines through particularly harsh irradiation environments suggest
that enhanced irradiation further drives the EW(5780)/EW(5797) ratio towards
lower values. This broadly corroborates variations noted by Welty et al.\
within each of their Magellanic Clouds and Milky Way samples. Likewise, Vos et
al.\ (2011) find a ratio towards the Upper Scorpius star-forming region
similar to that in our sightlines, $EW(5797)/EW(5780)\sim0.2$, but with
deviations in individual sightlines up to $EW(5797)/EW(5780)\sim1$.

%
% FIGURE 13
%
\begin{figure}
\centerline{\psfig{figure=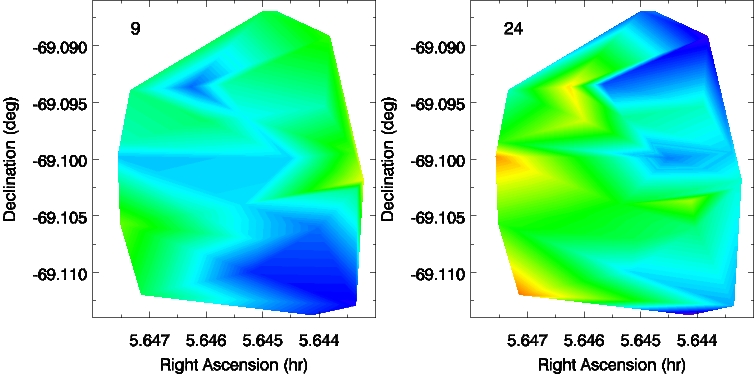,width=90mm}}
\caption[]{Velocity slices through the Galactic component of the Na\,{\sc
i}\,D$_1$ absorption in the UVES spectra, using a ``rainbow'' intensity scale
(with absorption increasing from purple/blue, through green/yellow to red; the
intensity scale has the same range in both panels but it is centered on the
median value in each panel). The velocity ranges are 7 and 15 km s$^{-1}$ for
the slices centered at 9 and 24 km s$^{-1}$, respectively.}
\label{sodium2}
\end{figure}

%
% TABLE 3
%
\begin{table}
\caption{FWHMs (in \AA) of the Galactic and LMC components of the 5780, 5797
and 6614 \AA\ DIBs: their median value, standard deviation and error in the
mean.}
\begin{tabular}{l|ccc|ccc}
\hline\hline
component & \multicolumn{3}{|c}{Galactic} & \multicolumn{3}{|c}{LMC} \\
 & media\rlap{n} & st.dev. & error & media\rlap{n} & st.dev. & error \\
\hline
5780 \AA\ DIB & 2.34 & 0.35 & 0.07 & 1.97 & 0.33  & 0.07 \\
5797 \AA\ DIB & 0.86 & 0.17 & 0.03 & 0.81 & 0.15  & 0.03 \\
6614 \AA\ DIB & 0.97 & 0.47 & 0.03 & 1.24 & 0.30  & 0.02 \\
\hline
\end{tabular}
\end{table}

Our LMC sample displays a ratio of equivalent widths of the 6614 and 5780 \AA\
DIBs, $EW(6614)/EW(5780)\sim0.2$ (up to $\sim0.4$ in some cases). This
contrasts with the Galactic data presented in Friedman et al.\ (2011) which
suggests that at a given strength of the 5780 \AA\ DIB, the LMC sightlines
have 5797 and 6614 \AA\ DIBs that are only half as strong. But this ratio
varies in the Milky Way, too, with $EW(6614)/EW(5780)\simeq1$ in the sample
presented by Moutou et al.\ 1999), but only $EW(6614)/EW(5780)\sim0.4$ towards
the Upper Scorpius star-forming region (Vos et al.\ 2011). We thus conclude,
for now, that our LMC sightlines show either enhanced 5780 \AA\ DIBs or
depleted 5797 and 6614 \AA\ DIBs, by a factor of two at most, or a combination
of both (by smaller factors).

%
% FIGURE 14
%
\begin{figure}
\centerline{\vbox{
\psfig{figure=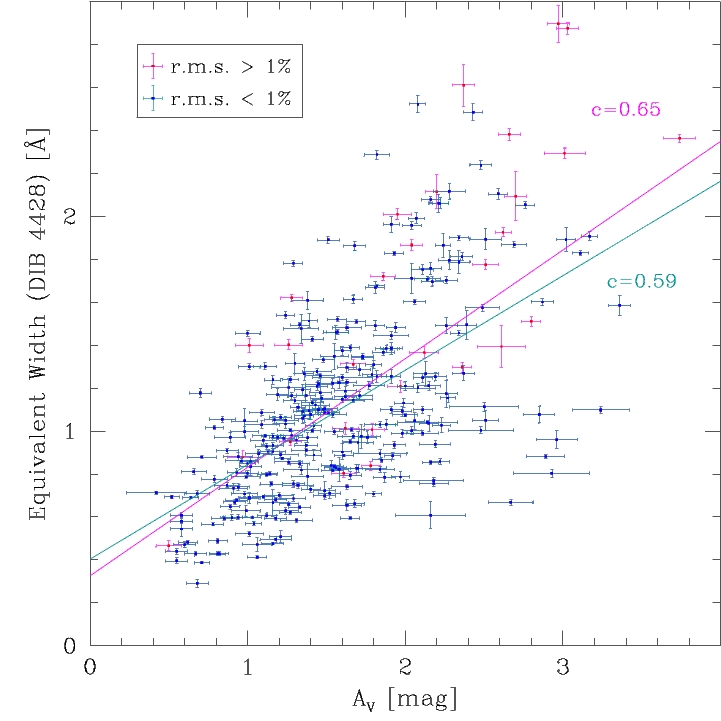,width=90mm}
\psfig{figure=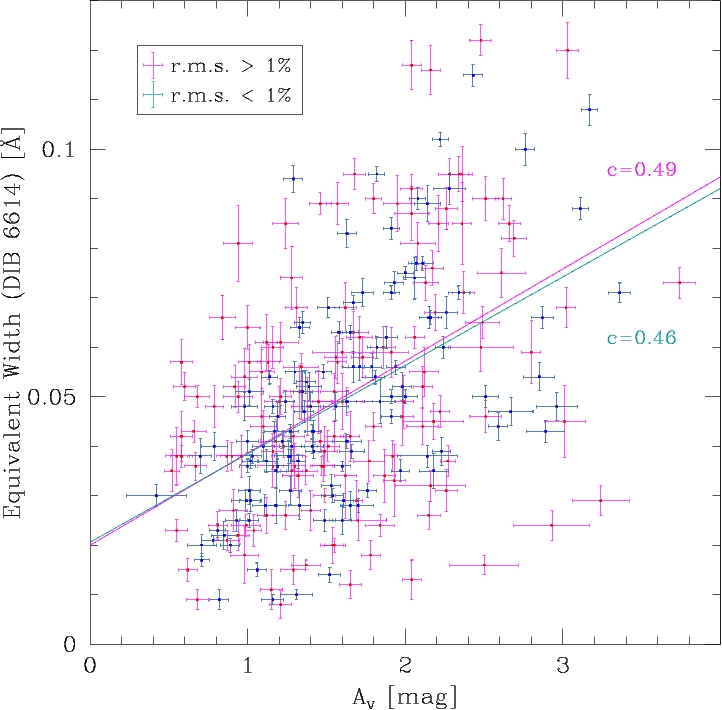,width=90mm}
}}
\caption[]{Correlations between the ({\it Top:}) 4428 \AA\ and ({\it Bottom:})
6614 \AA\ DIBs (LMC + Galactic), and broad-band visual extinction (Ma\'{\i}z
Apell\'aniz et al.\ in prep.). Symbols as in Fig.\ 9.}
\label{correlation3}
\end{figure}

%
% FIGURE 15
%
\begin{figure}
\centerline{\psfig{figure=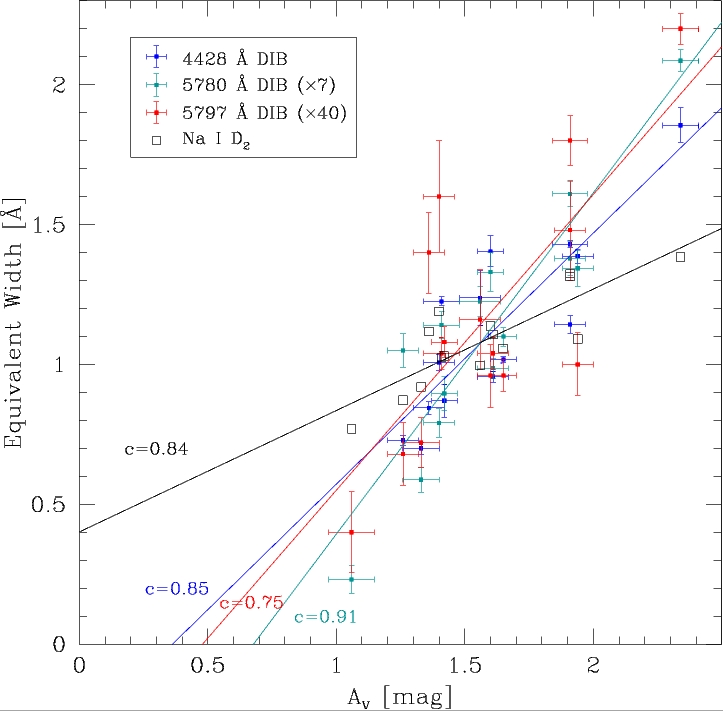,width=90mm}}
\caption[]{Correlations between the 4428 \AA\ DIB (blue), 5780 \AA\ DIB
(cyan), 5797 \AA\ DIB (red) and Na\,{\sc i}\,D$_2$ (black), and broad-band
visual extinction (Ma\'{\i}z Apell\'aniz et al.\ in prep.) for sightlines in
common. Linear regression lines are overplotted, and annotated with the linear
correlation coefficient.}
\label{correlation3}
\end{figure}

We now look at how the DIBs compare to the attenuation by dust grains,
realising that we cannot separate the Magellanic dust from the Galactic dust
though it appears that both the dust and DIBs are dominated by the LMC
component. We find $EW(5780)/A_{\rm V}\sim0.12$ \AA\ mag$^{-1}$ around $A_{\rm
V}\sim2$ mag, albeit with a possible threshold in A$_{\rm V}$. This is
consistent with the data in Welty et al.\ (2006), which suggest
$EW(5780)/A_{\rm V}=0.20$, 0.10 and 0.07 \AA\ mag$^{-1}$ for Galactic, LMC and
SMC sightlines, respectively (assuming $R_{\rm V}=A_{\rm V}/E(B-V)=3.1$);
Raimond et al.\ (2012) and Vos et al.\ (2011) find somewhat lower Galactic
values of $EW(5780)/A_{\rm V}=0.17$ and 0.15 \AA\ mag$^{-1}$, respectively. The
data presented by Vos et al.\ (2011) also suggest $EW(6614)/A_{\rm V}=0.06$
\AA\ mag$^{-1}$ which is double the rate that we find. Given that the Upper
Scorpius star-forming region studied by Vos et al.\ appears to exhibit a
similar radiation environment, as traced by the EW(5797)/EW(5780) and
EW(5780)/A$_{\rm V}$ ratio, and assuming that the visual extinction is roughly
proportional to metallicity, the simplest conclusion is that the 6614 \AA\ DIB
is disproportionally weaker in the Tarantula Nebula.

Snow et al.\ (2002b) observed a levelling off of the strength of the 4428 \AA\
DIB for $A_{\rm V}>3$ mag, of which there is a hint also in our data; up to
that point, $EW(4428)/A_{\rm V}\sim0.7$ \AA/mag both in their and our datasets.
Indeed, we noted sightlines with lots of dust but weak DIBs. Looking at where
these are located on the sky (Fig.\ 18) it appears that the smallest
$EW(4428)/A_{\rm V}$ ratios are found towards the immediate south--west and
north--east of R\,136 -- which incidentally is also where the far-IR dust
emission is most intense (Fig.\ 16). As the visual extinction is not low the
stellar probes are clearly {\it not} located in front of all of that dust and
the weak DIBs imply their carriers are not directly associated with the large
grains that dominate the far-IR emission. Snow et al.\ (2002a) found unusually
weak DIBs (at a given $E(B-V)$) in HD\,62542, and offered the explanation that
this might result from the absence of significant diffuse cloud envelopes (in
which the DIBs would be presumed to reside). Possibly, the eroding effect of
the O stars in R\,136 has taken its toll on the presence of all DIBs, not just
the 6614 \AA\ DIB (see above) but even the resilient 4428 \AA\ DIB. However,
if the flattening with A$_{\rm V}$ is related to a similar flattening with
EW(5780) noted in Section 3.2, then the explanation is more likely to involve
the saturation of the formation of the carrier of the 4428 \AA\ DIB or its
transformation into another DIB carrier or incorporation into dust, rather
than a more destructive scenario.

%
% FIGURE 16
%
\begin{figure*}
\centerline{\psfig{figure=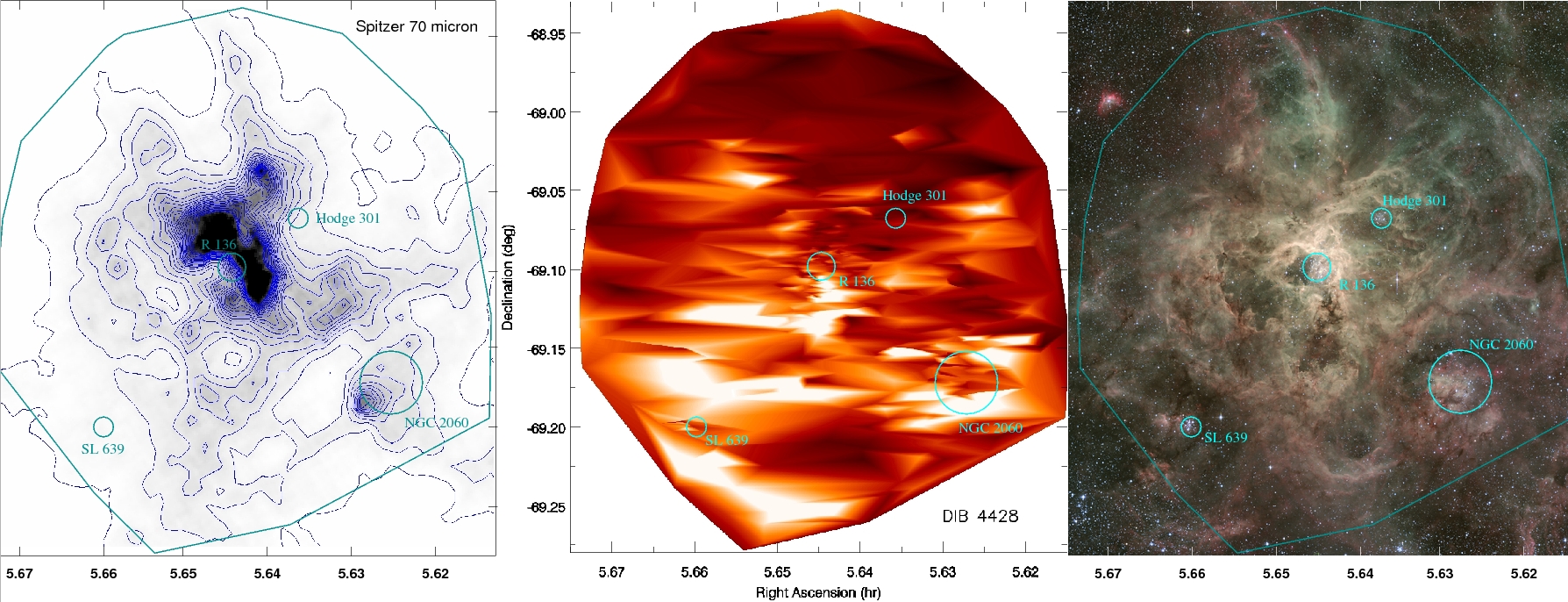,width=184mm}}
\caption[]{Comparison between ({\it Center:}) the map of the total equivalent
width of the 4428 \AA\ DIB, and ({\it Left:}) the {\it Spitzer} 70-$\mu$m
image of dust emission (Meixner et al.\ 2006) and ({\it Right:}) an optical
composite where blue is the B-band, green is the V-band plus [O\,{\sc iii}]
and red is H$\alpha$ (credit: ESO). The four most conspicuous star clusters
are marked and labelled.}
\label{overview}
\end{figure*}

%
% FIGURE 17
%
\begin{figure}
\centerline{\psfig{figure=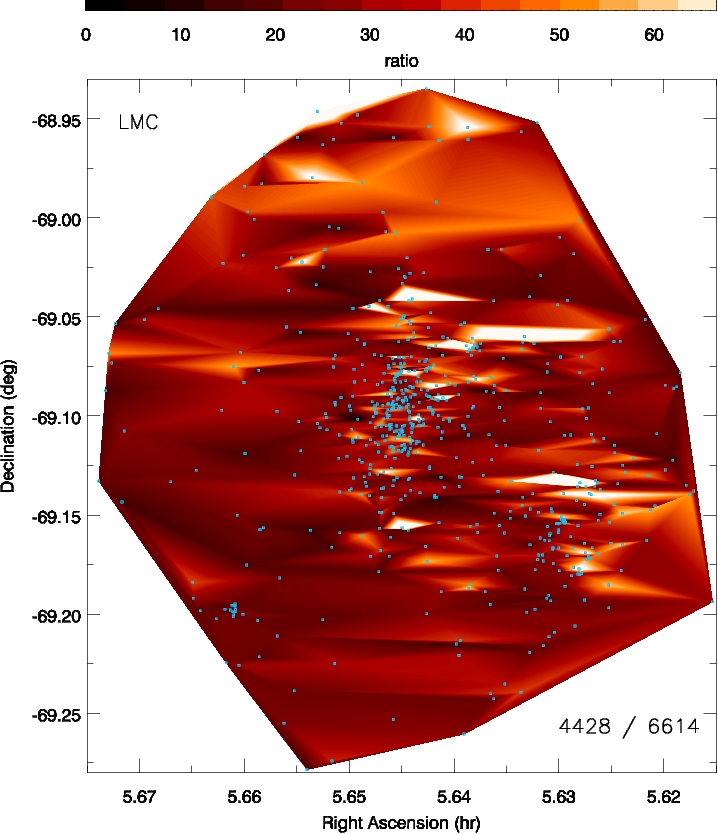,width=90mm}}
\caption[]{Ratio map of equivalent widths of the 4428 and 6614 \AA\ DIBs (LMC
component). The sightlines are marked with little blue dots.}
\label{correlationmap1}
\end{figure}

%
% FIGURE 18
%
\begin{figure}
\centerline{\psfig{figure=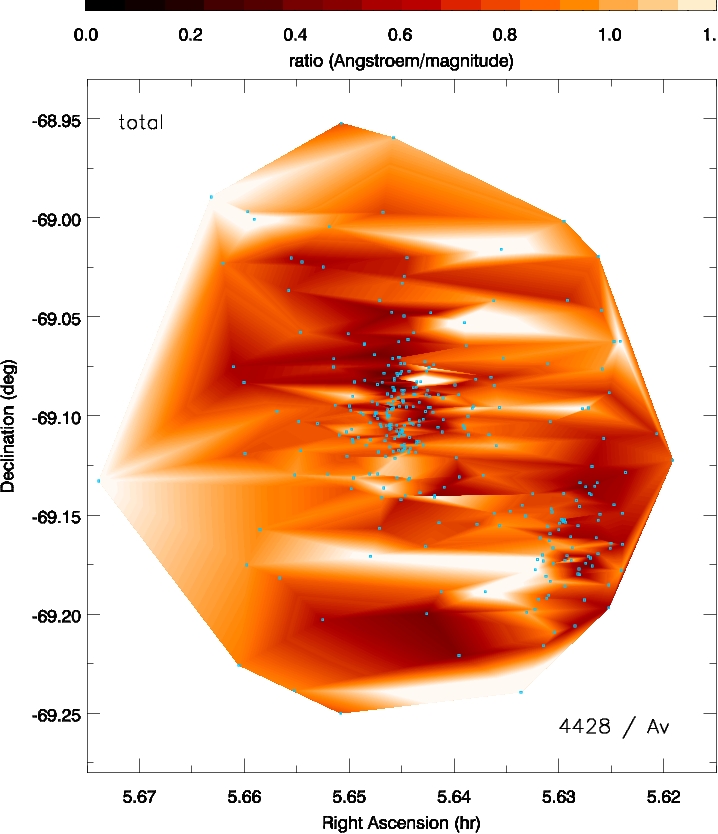,width=90mm}}
\caption[]{Ratio map of the (total) equivalent width of the 4428 \AA\ DIB and
the visual extinction. The sightlines are marked with little blue dots.}
\label{correlationmap2}
\end{figure}

While we have noticed that the Galactic sightlines in our spectra resemble
those of the LMC in terms of radiation environments -- for instance in terms
of EW(5797)/EW(5780) -- this is not to say that the high Galactic latitude gas
is similar to that in the LMC. Comparing the H\,{\sc i} (21-cm spin-flip
transition, McClure-Griffiths et al.\ 2009), Na\,{\sc i}\,D and 5797 \AA\ DIB
(Fig.\ 19), we find that the H\,{\sc i} is $\sim4$ times more intense in
30\,Dor than in the foreground gas, but the Na\,{\sc i}\,D absorption is about
equally strong suggesting a metal depletion of a factor four. The 5797 \AA\
DIB is about three times stronger in 30\,Dor than in the Galactic foreground,
suggesting a rather high DIB abundance given the lower metallicity of the LMC.
Note also in this comparison that the non-negligible intrinsic width of the
5797 \AA\ DIB (see above) means that its profile is narrower than that of the
Na\,{\sc i}\,D line and so not much of it can be associated with the clouds
that deviate kinematically from the bulk of the 30\,Dor ISM. It thus appears
that the carrier of the 5797 \AA\ DIB is more prevalent in the generally
denser and more neutral ambient ISM and less so in the more energetic
components, in contrast to the carrier of the 5780 \AA\ DIB, thus driving the
variations of the EW(5797)/EW(5780) ratio.

%
% FIGURE 19
%
\begin{figure}
\centerline{\psfig{figure=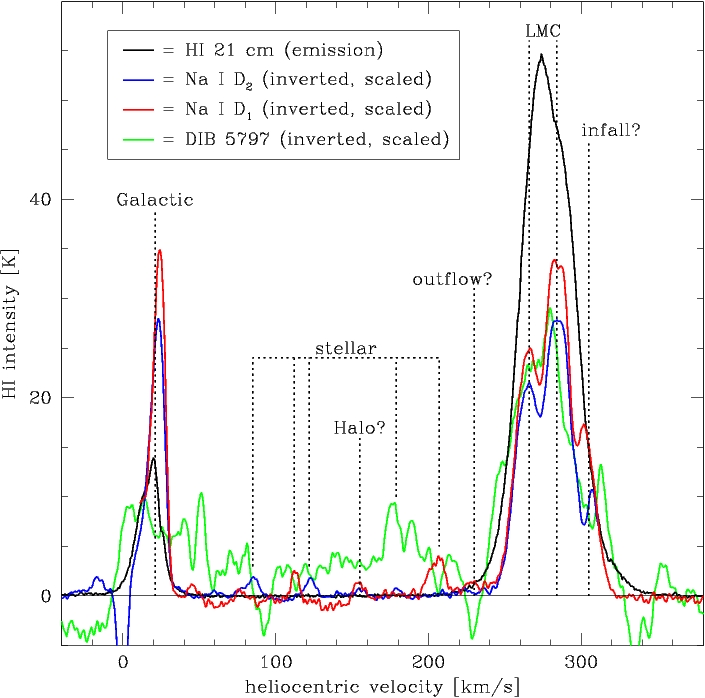,width=90mm}}
\caption[]{Line profiles of ({\it black:}) H\,{\sc i} emission
(McClure-Griffiths et al.\ 2009), ({\it blue:}) Na\,{\sc i}\,D$_2$, ({\it
red:}) Na\,{\sc i}\,D$_1$ and ({\it green:}) 5797 \AA\ DIB. The Na and
(lightly smoothed) DIB profiles are inverted and arbitrarily scaled for ease
of comparison. Pertinent features are marked and labelled.}
\label{comparison}
\end{figure}

Friedman et al.\ (2011) observed a generally good correlation between DIB
strength and H\,{\sc i} column density. While the H\,{\sc i} emission would
indeed serve as a useful reference against which to gauge the strength of
other tracers of the ISM, the limited resolution of the available H\,{\sc i}
surveys of the LMC ($1^\prime$, Kim et al.\ 1998) at present precludes such
detailed correlation; we note that on these scales ($\sim15$ pc) the H\,{\sc
i} maps resemble the far-IR maps (Meixner et al.\ 2010) such as the one in
Fig.\ 16, and that a high-resolution, sensitive H\,{\sc i} survey of the
Magellanic Clouds is going to be carried out with the Australian Square
Kilometre Array Pathfinder (Dickey et al.\ 2012).

%....................................................................... 4.1.3
\subsubsection{Minor DIBs}

We now briefly consider additional DIBs in the spectral ranges covered by the
VFTS. We recall the detection of the 4502 \AA\ DIB in our LMC sightlines
(Fig.\ 1), and its strong presence in the Galactic massive young star cluster,
NGC\,3603, and in particular in the Galactic H\,{\sc ii} region RCW\,49 where
it is about half as deep as the 4428 \AA\ DIB (Morrell, Walborn \& Fitzpatrick
1991). Compared to the other DIBs we have described, the 4502 \AA\ DIB appears
stronger in the LMC than in our Galactic sightlines. These observations point
at UV irradiation as a driver behind the strong presence of the carrier of the
4502 \AA\ DIB, perhaps even more so than for that of the 4428 \AA\ DIB. We
also wish to draw the attention to a set of DIBs at 4727, 4762 and 4780 \AA,
covered with the LR03 setting of the Medusa mode (at a spectral resolution of
0.56 \AA); while too weak to study in detail in individual spectra, they are
clearly detected both in the LMC and Galactic foreground (Fig.\ 20).
Interestingly, the Galactic-to-LMC ratio of the 4727 \AA\ DIB seems relatively
low compared to that of the 4762 and 4780 \AA\ DIBs, indicating differences in
their carriers; there is thus a clear diagnostic value to consider these DIBs
jointly.

%
% FIGURE 20
%
\begin{figure}
\centerline{\psfig{figure=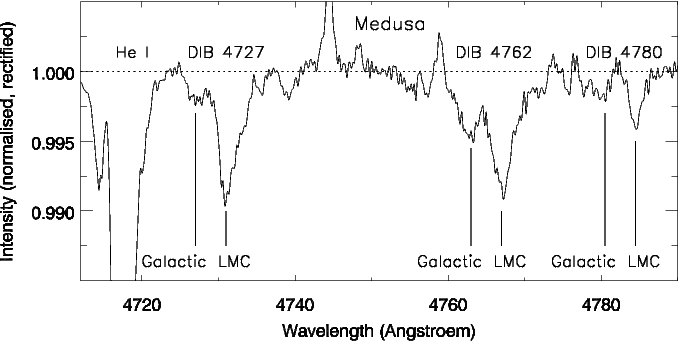,width=90mm}}
\caption[]{Normalised and rectified, r.m.s.(squared)-weighted average spectrum
of the 4727, 4762 and 4780 \AA\ DIBs in the Medusa sample.}
\label{correlationmap2}
\end{figure}

%------------------------------------------------------------------------- 4.2
\subsection{Structure and kinematics of the ISM}

Apart from mapping the locations and excitation environments of the unknown
carriers of the DIBs, we have kinematic information from the Na\,{\sc i}\,D
lines to help make sense of the ISM structures and the relation between DIBs
and atomic gas. Furthermore, we can quantify spatial variations of absorption
at small scales to investigate what underlying structures are inherent in the
maps and correlations that we discussed earlier.

%....................................................................... 4.2.1
\subsubsection{An expanding bubble around R\,136?}

Walborn (1980), Blades (1980) and Blades \& Meaburn (1980) observed R\,136 in
Na\,{\sc i}\,D and Ca\,{\sc ii}\,K absorption and found, besides one Galactic
component, four LMC components in the range 242--305 km s$^{-1}$ with a wing
extending to about 150 km s$^{-1}$. They noted large variations in equivalent
width between Na\,{\sc i}\,D and Ca\,{\sc ii}\,K; the latter was stronger in
the more extreme blue and red components, indicative of ionized gas or of the
Routly--Spitzer effect (Routly \& Spitzer 1952) explained by Barlow \& Silk
(1977) due to sputtering of calcium atoms off of dust grains. Both these
mechanisms, ionization and sputtering are associated with hot gas. It is
therefore not surprising that the 5797 \AA\ DIB does not appear to trace these
extreme kinematic components.

The massive cluster R\,136 clearly impacts upon the surrounding ISM by its
ionizing radiation and the ram pressure from its stellar winds; this is seen
in the form of a cavity devoid of dust and DIBs (Fig.\ 16), but also through
its imprint in the velocity maps of the Na\,{\sc i}\,D absorption (Fig.\ 12).
A high-velocity blob of absorbing sodium in the sightline towards VFTS\,482
suggests an expansion at a rate of 40 km  s$^{-1}$ which, for a shell of 6 pc
radius corresponds to a dynamical age of $<10^5$ yr, i.e.\ much less than the
age of R\,136 ($\sim10^6$ yr). This could imply a delay between the formation
of the cluster and the onset of a powerful cluster wind as the O-type stars
evolve off of the main sequence. The spectral line associated with this fast
blob has an equivalent width of $W\sim0.04$ \AA\ (cf.\ Table 2). This can be
converted to a column density in the lower level of the transition (see, e.g.,
van Dishoeck \& Black 1989):
\begin{equation}
N = 1.13\times10^{20}\frac{W}{f\lambda^2},
\end{equation}
where the oscillator strength $f=0.33$ for Na\,{\sc i}\,D$_1$. We thus obtain
$N\sim4\times10^{11}$ cm$^{-2}$. If we assume that this represents the majority
of sodium, that the sodium abundance in the LMC is half solar (i.e.\
$\log\epsilon_{\rm H}=6+\log\epsilon_{\rm Na}$, see Baum\"uller, Butler \&
Gehren 1998), and that hydrogen constitutes 75\% of the material in mass, then
we obtain a surface mass density of $\Sigma\sim9$ mg m$^{-2}$. For a uniform
shell this would amount to $4\pi R^2$ as much, where $R\sim6$ pc is the radius
of the shell, i.e.\ 2 M$_\odot$. For typical mass-loss rates of evolved O
stars of $10^{-6}$--$10^{-5}$ M$_\odot$ yr$^{-1}$ (Vink, de Koter \& Lamers
2001; Mokiem et al.\ 2007a,b) this amount could have been produced by a mere
handful of O stars over the course of $10^5$ yr. The winds will have swept up
some interstellar gas, though; if half of it was swept-up, then the average
ISM density within the 6-pc-radius bubble would have been $n\sim0.04$
cm$^{-3}$, i.e.\ typical of the warm, weakly-ionized ISM.

But Na\,{\sc i} is {\it not} the dominant ionization stage except for very
dense clouds, and it also is depleted within dust grains even in diffuse ISM.
So the above estimate for the mass in the shell will be too low. Wakker \&
Mathis (2000) showed that the Na\,{\sc i} abundance shows very little
dependence on H\,{\sc i} column density; it is lower than solar by a factor of
a few hundred, with scatter over two orders of magnitude. This would bring the
mass of the shell around R\,136 closer to $10^3$ M$_\odot$. As it would
require a thousand evolved O stars to supply this amount of matter within less
than $10^5$ yr it would be more likely that much -- if not most -- of the mass
is swept-up ISM. This would also be consistent with the expansion speed to
have dropped from over 1000 km s$^{-1}$ in the winds of the O stars down to
the 40 km s$^{-1}$ of the shell. The revised mass would correspond to a prior
average ISM density in the range of 10--100 cm$^{-3}$, more typical of cool
ISM perhaps related to the formation of R\,136.

That said, the blue-shifted absorption component is not seen in all of the
UVES spectra, and so the assumption of a uniform shell cannot be correct. This
would bring down the estimate of the total mass in the shell. On balance, we
conjecture that it is plausible that a number (dozens?) of evolved O stars
within R\,136 together have blown a bubble of 6 pc radius in less than $10^5$
yr, sweeping up a sizeable fraction of its mass from the local, cool but not
particularly dense ($n\sim1$ cm$^{-3}$?) ISM.

The picture may not be that simple, though. Firstly, gas may actually also be
{\it falling towards} 30\,Dor at $\sim40$ km s$^{-1}$ as well, as loosely
remarked by Blades (1980) and also visible in our velocity maps (Figs.\ 12, 19
\& 21). This must be interpreted in light of the paradoxial scenarios of gas
outflow (Nidever, Majewski \& Burton 2008) and infall (Olsen et al.\ 2011) as
deduced from H\,{\sc i} emission and possibly associated stellar kinematics,
respectively. The 30\,Dor region at large may well be accreting cool gas
whilst ejecting hot gas, much akin the ``fountains'' of the Milky Way Disc
(see, e.g., Marasco, Fraternali \& Binney 2012, and references therein).

Secondly, while expulsion by stellar winds is a possibility the blue-shifted
(w.r.t.\ the systemic velocity of $\sim275$ km s$^{-1}$) gas may reside
instead in the Milky Way Halo (Blades 1980; Blades \& Meaburn 1980). De Boer,
Koornneef \& Savage (1980) found UV absorption at 20 km s$^{-1}$ (Galactic)
and at 220, 250 and 290 km s$^{-1}$ towards the LMC and proposed that the 220
km s$^{-1}$ component might originate in coronal gas associated with the LMC.
Intermediate- and high-velocity clouds floating in the Galactic Halo have also
been seen in front of the LMC at 60 and 120 km s$^{-1}$, interestingly also in
H$_2$ absorption (Richter et al.\ 1999; Richter, Sembach \& Howk 2003; Lehner,
Staveley-Smith \& Howk 2009). We only, tentatively, identify one kinematic
component between 30 and 230 km s$^{-1}$, viz.\ at $\sim155$ km s$^{-1}$ (Fig.\
19); other absorption components are not consistently present in both Na\,{\sc
i}\,D$_1$ and D$_2$ components and must thus be stellar (i.e.\ not sodium) in
origin. The 155 km s$^{-1}$ component might be associated with a Halo cloud,
perhaps originating in the LMC (Lehner et al.\ 2009) and also seen in more
sensitive H\,{\sc i} observations (cf.\ McGee, Newton \& Morton 1983).
However, there is no sign of H\,{\sc i} emission at 155 km s$^{-1}$ in the
direction of R\,136 (Fig.\ 19).

%....................................................................... 4.2.2
\subsubsection{Small scale structure of the ISM}

On the scale of the Medusa maps (Figs.\ 4 \& 6) -- viz.\ $20^\prime$ or 300 pc
at the distance of the LMC and $<0.6$ pc for gas within 100 pc distance -- the
DIB absorption varies considerably; variations are still seen at even smaller
scales (Figs.\ 5, 7 \& 8) also in Na\,{\sc i} (Figs.\ 12 \& 13). We quantify
these variations by the standard deviation and also express these in terms of
the median value (Table 4). The latter are typically between 0.1 and 1, which
is expected for a medium with structure across a range of scales -- the rather
high values of $\sim0.5$ are most readily explained by fairly short gas
columns, closer to $\sim0.1$ kpc than kpc lengths (van Loon et al.\ 2009). One
should bear in mind, of course, that if stars and gas are mixed then part of
the variance in absorption is due to the fact that some stars will be situated
toward the back of the entire gas column and others toward the front; this
cannot be quantified without adopting a model for the distribution of stars
and gas. If gas and stars share an identical, uniform distribution in space,
then it is trivial to derive that the standard deviation will be $\sim57$\% of
the median. More trivial still, if the stars are all behind all of the gas
then the standard deviation due to the mixing will be zero. The values in
Table 4 below 57\% could therefore be taken to indicate that most of the gas
probed by the maps is in front of the stars; one could imagine a sheet lying
in front of the clustered massive stars, for instance, consistent with the
idea of an expanding shell surrounding R\,136.

%
% FIGURE 21
%
\begin{figure}
\centerline{\psfig{figure=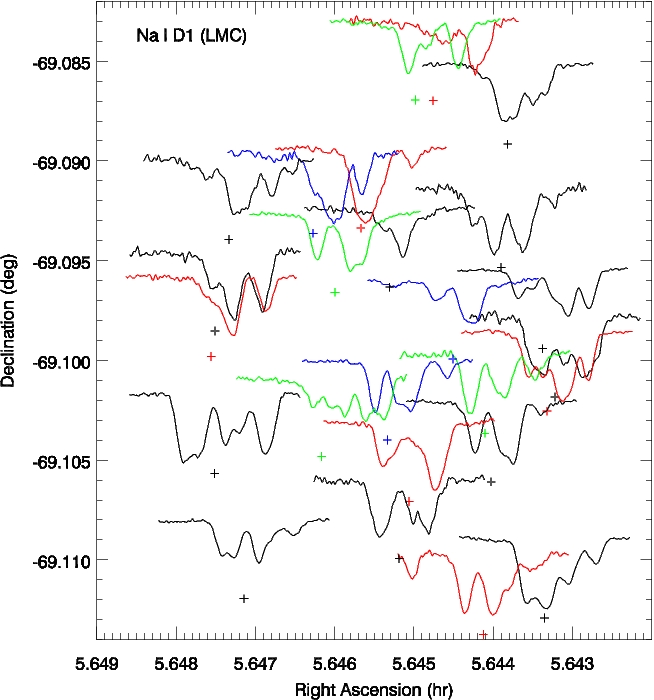,width=90mm}}
\caption[]{Na\,{\sc i}\,D$_1$ spectra of the LMC component, positioned on the
sky. Each spectrum covers 220--330 km s$^{-1}$; a plus marks the star's
position on the sky and the zero intensity and 275 km s$^{-1}$ point in the
spectrum.}
\label{spectramap}
\end{figure}

%
% TABLE 4
%
\begin{table*}
\caption{Summary of variations in absorption across $20^\prime$ (Medusa
spectra), $1\rlap{.}^\prime3$ (UVES spectra) and $0\rlap{.}^\prime5$ (ARGUS
spectra): the standard deviation (in \AA) and also in relation to the median
(in per cent) -- unless the latter is zero. The linear scales refer to a
canonical distance of 100 pc (Galactic) or the distance to the LMC (50 kpc),
respectively.}
\begin{tabular}{l|lrlrlr|lrlrlr}
\hline\hline
component      & \multicolumn{6}{|c}{Galactic}           &
                 \multicolumn{6}{|c}{LMC}                \\
               & \multicolumn{2}{c}{0.6 pc}              &
                 \multicolumn{2}{c}{0.04 pc}             &
                 \multicolumn{2}{c}{0.015 pc}            &
                 \multicolumn{2}{|c}{300 pc}             &
                 \multicolumn{2}{c}{20 pc}               &
                 \multicolumn{2}{c}{8 pc}                \\
               & \AA   &  \% & \AA   &  \% & \AA   &  \% &
                 \AA   &  \% & \AA   &  \% & \AA   &  \% \\
\hline
4428 \AA\ DIB  & 0.12  &     & 0.19  &     & 0.20  & 389 &
                 0.49  &  54 & 0.32  &  31 & 0.36  &  33 \\
5780 \AA\ DIB  &       &     & 0.015 &  34 &       &     &
                       &     & 0.070 &  56 &       &     \\
5797 \AA\ DIB  &       &     & 0.006 &  71 &       &     &
                       &     & 0.014 &  77 &       &     \\
6614 \AA\ DIB  & 0.008 &  86 &       &     &       &     &
                 0.026 &  79 &       &     &       &     \\
Na\,{\sc i}\,D &       &     & 0.013 &   7 &       &     &
                       &     & 0.11  &  25 &       &     \\
\hline
\end{tabular}
\end{table*}

Considering the angular scale of the UVES spectra, we see consistently larger
variations in the LMC than in the Galactic component, both in absolute terms
and relative to the median level. On scales from 0.04 pc to 20 pc, the
structures probed by DIBs and Na\,{\sc i} may be losing some of their
coherency. This idea is corroborated by the larger variation in the Galactic
6614 \AA\ DIB on scales of 0.6 pc as compared to the variations on a scale of
0.04 pc; likewise, the LMC 6614 \AA\ DIB and 4428 \AA\ DIB vary more on even
larger scales, of $\sim300$ pc. (Note that the measurements of the Galactic
4428 \AA\ DIB component are very dubious.) This all suggests that the ISM
probed by DIBs and Na\,{\sc i}, i.e.\ neutral and weakly-ionized gas, has
structure on scales ranging from as small as $\sim0.1$ pc to as large as
$\sim100$ pc -- spanning three orders of magnitude. The variations are larger
for the 5797 \AA\ DIB than for the 5780 \AA\ DIB, which are larger than for
the Na\,{\sc i} (both in the Galactic and LMC components). This suggests that
there is more, smaller structure in the colder ISM phase (e.g., probed by the
5797 \AA\ DIB) than in the warmer ISM phase (e.g., probed by the 5780 \AA\ DIB
and especially the Na\,{\sc i} line). Such explanation may also apply to the
larger variations in the LMC 6614 \AA\ DIB as compared to the LMC 4428 \AA\
DIB as the latter is thought to exist (also) in warmer gas. This picture is
also consistent with that proposed by Pan et al.\ (2005), in which tracers of
warmer ISM are distributed more widely in space as well as kinematically
whilst tracers of colder ISM are confined to dense regions and generally show
narrow line profiles. These differences give rise to scatter in -- and
deviations from -- correlations that in essence trace the overall gas column
density. A dependence of structure formation on gas temperature such as
outlined here is expected as structure develops when cooling gas condenses
and/or fragments due to thermal instabilities. The 100-pc scales may be
related to molecular cloud complexes and inter-cloud regions, whilst pc and
sub-pc scales may be related to cloud cores, stellar wind bubbles and
structure in cool--warm ISM interfaces.

The Na\,{\sc i} line profiles (Fig.\ 21) show very significant variation in
several discrete absorption components (clouds) on scales as little as
$4^{\prime\prime}$ (1 pc) in the LMC. When cross-correlating these spectra (Fig.\
22) we see a gradual increase in their differences as their angular separation
increases. This is true both for the Galactic and LMC component, implying
structure (in the Galactic gas, assuming a distance of 100 pc) on scales as
small as $\sim0.01$ pc (just 2000 AU). The latter could correspond to the
interstellar cirrus filaments that are sometimes visible in optical or far-IR
images of extra-planar gas. Tiny- and small-scale structure appears to be
commonplace in a wide variety of environments within our Galaxy -- see Smoker
et al.\ (2011) and references therein. Pan et al.\ (2005) combine
spectroscopic observations with a chemical model to relate spectral variations
seen towards $\rho$\,Oph to column density variations up to a factor ten on
scales of $10^4$ AU, in addition to the typical pc-scale variations seen for
instance across the Cepheus bubble in CO. They suggest the high contrast
implies sheet-like structures as proposed by Heiles (1997), with aspect ratios
of 5--10. While we clearly notice tiny-scale structure in our data the
amplitude is below unity and therefore does not require sheet-like or
filamentary cloud geometry; the drastic variations in individual kinematic
components that we see in our LMC sightlines are related to the wider
distribution of individual clouds, on scales exceeding a parcec. Complex,
sub-pc structure has also been seen towards the hyper-massive eruptive star
$\eta$\,Carinae and stars in the adjacent Trumpler\,16 cluster in the Milky
Way, which have been linked to possible outflows from a supernova remnant or
protostellar outflows in front of the stars (Walborn et al.\ 2007).

%
% FIGURE 22
%
\begin{figure}
\centerline{\psfig{figure=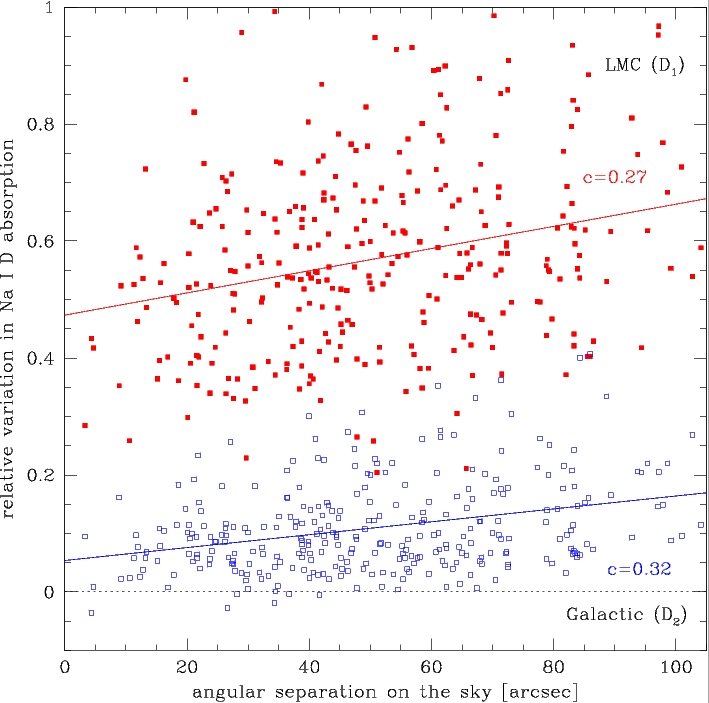,width=90mm}}
\caption[]{Relative variation in Na\,{\sc i}\,D absorption as a function of
angular separation on the sky, for the ({\it red solid:}) D$_1$ line of the
LMC component and the ({\it blue open:}) D$_2$ line of the Galactic component.
Linear regression lines are overplotted, and annotated with the linear
correlation coefficient.}
\label{coherence}
\end{figure}

Small-scale structure on sub-pc scales in the local (cold) ISM towards the
Magellanic Clouds has been found recently also in UV data. Nasoudi-Shoar et
al.\ (2010) analyzed molecular hydrogen (H$_2$) absorption in the local
foreground gas along several closely spaced sightlines towards early-type
stars in the N\,51\,D superbubble in the LMC. Based on the analysis of the
spatially varying H$_2$ absorption pattern and the physical conditions in the
gas, these authors conclude that the local cold neutral medium towards the LMC
consists of small-scale structure on linear scales $<0.1$ pc. These results
are in good agreement what is seen in our Na\,{\sc i}\,D data towards 30\,Dor,
which is not surprising as Na\,{\sc i} and H$_2$ are expected to trace a
similar gas phase in the Milky Way Disc.

%=========================================================================== 5
\section{Conclusions}

We have used over 800 sightlines probed by spectra from the VFTS covering the
Tarantula Nebula, to measure the absorption by the DIBs at 4428 and 6614 \AA,
and 25 sightlines in a more restricted region centered on the central cluster
R\,136, to measure the absorption by the DIBs at 5780 and 5797 \AA\ and by the
Na\,{\sc i}\,D doublet. We have used these measurements to create maps of the
4428 and 6614 \AA\ DIBs and maps of the Na\,{\sc i}\,D absorption in different
velocity slices. We have compared the absorption probed by these spectral
features, and with interstellar reddening. Our main conclusions can be
summarised as follows:
\begin{itemize}
\item[$\bullet$]{The maps of the 4428 \AA\ and 6614 \AA\ DIBs bear little
resemblance to images of the far-IR emission from cool dust, indicating that
the carriers of DIBs are not directly related to relatively large grains (in
keeping with an observed break-down of the correlation between DIB strength
and visual extinction for large values of the latter); the features in the DIB
maps more closely correspond to nebulosity traced by hydrogen recombination
(H$\alpha$) and tend to avoid superbubbles traced by [O\,{\sc iii}] emission
and the immediate vicinity of OB associations.}
\item[$\bullet$]{The molecular cloud complex to the south of 30\,Doradus is
traced -- at least in a general sense -- by the carriers of the 4428 and 6614
\AA\ DIBs, but the carrier of the 4428 \AA\ DIB is also present in warmer
diffuse gas found throughout the Tarantula Nebula.}
\item[$\bullet$]{Differences exist in the relationships between the different
DIBs and Na, with the 4428 \AA\ DIB present already at low Na column density
but the 6614, 5780 and 5797 \AA\ DIBs only starting to appear through
subsequently denser Na columns; a similar trend is observed for the relation
between the DIBs and visual extinction in common sightlines close to R\,136.
In agreement with the previous conclusions, we suggest that the carrier of the
4428 \AA\ DIB may be a relatively large yet compact, and electrically neutral
molecule that is relatively resilient to impacting energetic photons or
particles, as opposed to a relatively small, fragile molecular carrier of the
5797 \AA\ DIB; the carriers of the 6614 and 5780 \AA\ DIBs may have
intermediate sizes and they may be mildly electrically charged, making them
vulnerable to destruction in exposed environments and to recombination within
dense clouds -- cf.\ Sandstrom et al.\ (2012) for similar considerations with
regard to the carriers of unidentified IR bands in the Magellanic Clouds.}
\item[$\bullet$]{The 6614, 5780, 5797 and especially 4428 \AA\ DIBs correlate
with the extinction of the (same) star by interstellar grains. In light of the
poor correlation of DIBs with dust emission this suggests that moderate visual
extinction may be dominated by relatively small grains that are mixed with
warm neutral or weakly-ionized gas, rather than the large grains that dominate
far-IR emission.}
\item[$\bullet$]{Both the Tarantula Nebula and Galactic high-latitude gas are
characterised by strong radiation fields, as evinced by the weak 5797 \AA\ DIB
compared to the 5780 \AA\ DIB. Differences are found between the Galactic and
LMC sightlines in the ratios of the 4727 \AA\ DIB and 4762 and 4780 \AA\ DIBs,
however, which could make these DIBs valuable additional diagnostics --
perhaps of metallicity.}
\item[$\bullet$]{Compared to Galactic samples, at a given visual extinction
the 5780 \AA\ DIB is weaker while the 4428 \AA\ DIB is of similar strength.
Assuming the extinction scales in proportion to metallicity this would suggest
that the abundance of the carrier of the 4428 \AA\ DIB is proportional to
metallicity and less dependent on radiation field while the other DIBs are
diminished by metallicity as well as irradiation. The same conclusion is
reached when Na\,{\sc i} is used as reference instead of extinction.}
\item[$\bullet$]{Stellar winds from the central cluster R\,136 appear to have
created a shell of $\sim10^2$ M$_\odot$ which is expanding at a speed of
$\sim40$ km s$^{-1}$. Some evidence for infall is also present, suggesting the
operation of a galactic ``fountain''.}
\item[$\bullet$]{Structure is detected in the distribution of cool--warm gas
down to scales of a few pc in the LMC and as little as $0.01$ pc in the Sun's
vicinity, more strongly so in the DIBs than in Na\,{\sc i}. This corroborates
the notion that the carriers of the DIBs reside in neutral or weakly-ionized
gas but not in hotter gas which is also traced by Na\,{\sc i}.}
\end{itemize}

%==============================================================================
\begin{acknowledgements}
We would like to thank the anonymous referee and Keith Smith for their useful
suggestions. AEB and BLT acknowledge STFC studentships awarded to Keele
University. JMA acknowledges support from the Spanish Government Ministerio de
Educaci\'on y Ciencia through grants AYA2010-15081 and AYA2010-17631, and from
the Consejer\'{\i}a de Educaci\'on of the Junta de Andaluc\'{\i}a through
grant P08-TIC-4075. STScI is operated by AURA, Inc.\ under NASA contract NAS
5-26555. This paper makes use of spectra obtained as part of the VLT--FLAMES
Tarantula Survey (ESO programme 182.D-0222). Based on observations made with
ESO Telescopes at the La Silla Observatory under programme ID 076.C-0888,
processed and released by the ESO VOS/ADP group.
\end{acknowledgements}

%========================================================================== A1
\appendix

\section{Line profile fitting}

The constraints on the fits to the DIBs are summarised in Table A1. The 4428
\AA\ DIB was best fit with Lorentzian profiles, to account for the broad
wings, whilst the 5780 and 5797 \AA\ DIBs and the 6614 \AA\ DIB were better
fit with Gaussian profiles, corroborating the findings by Snow et al.\ (2002b)
for the 4428 and 6614 \AA\ DIBs.

The 4428 \AA\ DIB is troubled by He\,{\sc i} lines at either side (at rest
wavelengths of 4387 and 4437 \AA, respectively), limiting the wavelength range
within which we fit the spectrum and thereby affecting the accuracy of the
continuum level determination. Also, sharp spectral lines occur in many B-type
spectra close to the centroid of the 4428 \AA\ DIB; we simultaneously fit the
three strongest of these lines using Gaussians profiles (they are identified
with red-shifted O\,{\sc ii} lines at rest wavelengths of 4415 and 4417 \AA,
and a red-shifted He\,{\sc i} line at a rest wavelength of 4437 \AA,
respectively). Other lines could potentially overlap with the DIBs -- e.g.,
the N\,{\sc ii} at a rest wavelength of 6610 \AA\ could interfere with the
Galactic 6614 \AA\ DIB; note also that binary motion could cause line shifts
that differ from the systematic velocity of the LMC. However, this was found
by visual inspection of all the spectra not to be a major problem.

We decided to fix the position of the Galactic and LMC components of the 4428
\AA\ DIB to $\lambda_{\rm Gal}=4428.5$ \AA\ and $\lambda_{\rm LMC}=4432.5$ \AA,
and its FWHM to 20 \AA, as this dramatically improved the reliability of the
fits; this was judged from the much closer resemblance between the 4428 and
6614 \AA\ maps of the LMC component as well as the better agreement between
the three stars in common between the Medusa and ARGUS spectra (especially
evident for the Galactic component). By doing this, the equivalent width of
the 4428 \AA\ DIB has become a proxy for its peak depth. The equivalent width
measurements of the DIBs are listed in Tables A2--A4, which are made available
electronically at the Centre de Donn\'ees astronomiques de Strasbourg (CDS).

%
% TABLE A1
%
\begin{table}
\caption{Fit constraints.}
\begin{tabular}{lllll}
\hline\hline
DIB                      &
function                 &
$\lambda_{\rm Gal}$ (\AA) &
$\lambda_{\rm LMC}$ (\AA) &
\llap{F}WHM (\AA\rlap{)} \\
\hline
4428                     &
Lorentzia\rlap{n}        &
4428.5 (fixed)           &
4432.5 (fixed)           &
20 (fixed)               \\
5780                     &
Gaussian                 &
5781.1--5781.9           &
5785.5--5786.3           &
1.32--2.45               \\
5797                     &
Gaussian                 &
5797.3--5797.7           &
5802.3--5802.7           &
0.75--1.13               \\
6614                     &
Gaussian                 &
6613.4--6614.4           &
6619.4--6620.4           &
0.59--1.77               \\
\hline
\end{tabular}
\end{table}

%
% FIGURE A1
%
\begin{figure}
\centerline{\psfig{figure=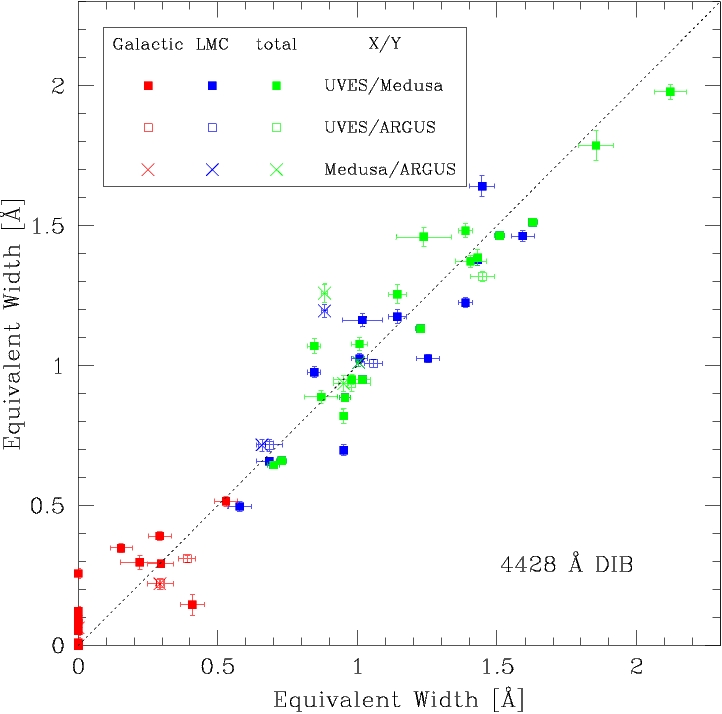,width=90mm}}
\caption[]{Equivalent width of the 4428 \AA\ DIB measured for common
sightlines among the UVES, Medusa and ARGUS spectra. The 1:1 correlation is
drawn as a dotted line for guidance.}
\label{common}
\end{figure}

%
% FIGURE A2
%
\begin{figure*}
\centerline{\hbox{
\psfig{figure=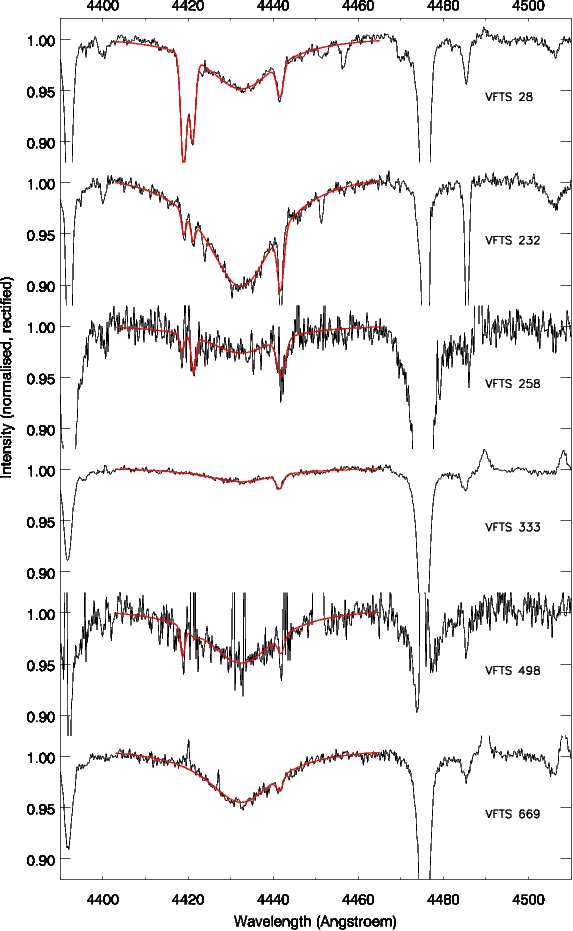,width=91mm}
\hspace{1mm}
\psfig{figure=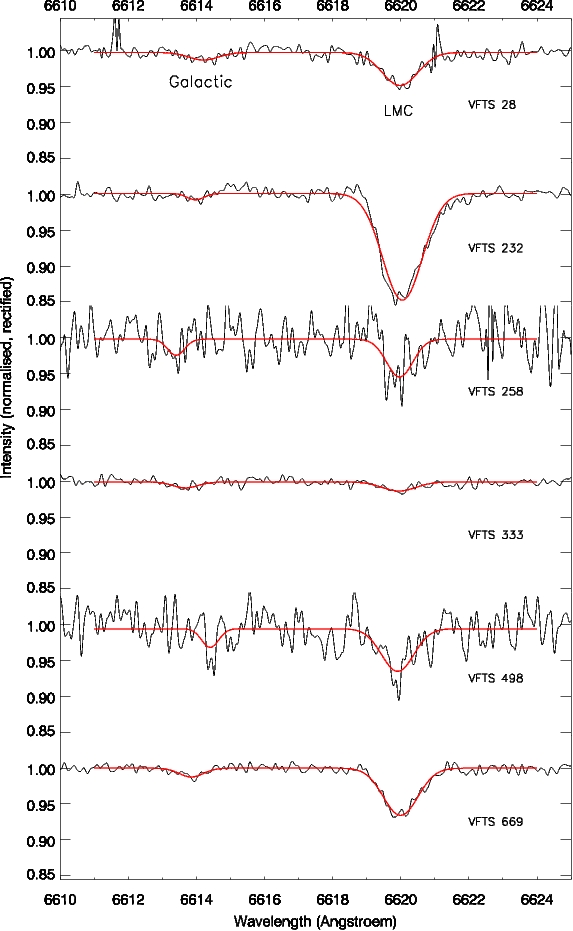,width=91mm}
}}
\caption[]{Examples of both high and poor signal-to-noise Medusa spectra and
different spectral complications, around the ({\it Left:}) 4428 \AA\ DIB and
({\it Right:}) 6614 \AA\ DIB, showcasing also the strongest DIBs, seen in the
direction of VFTS\,232.}
\label{common}
\end{figure*}

What constitutes a ``reasonable'' fit is somewhat arbitrary. However, when
selecting on the basis of the r.m.s.\ value in the normalisation intervals it
became clear upon inspection of the spectra and the fits that spectra with
r.m.s.\ $>3$\% almost never showed convincing fits except in rare cases of
very strong absorption. Vice versa, most spectra with r.m.s.\ $<3$\% showed
believable fits to absorption features at the expected places. Visual
inspection of all fits was necessary to remove a small number of fits that
were clearly affected by spectral artefacts or very strong emission lines (a
clipping procedure limiting the fitted data to within $-20$\% and +3\% of the
continuum level rendered the fitting already insensitive to sharp spikes).
This yielded a 98\% success rate for the 4428 \AA\ DIB in the Medusa spectra,
95\% in the ARGUS spectra and 96\% in the UVES spectra; the Galactic and LMC
components of the 6614 \AA\ DIB in the Medusa spectra were fit at a 76\% and
82\% success rate, respectively. The values in Tables A2--A4 are flagged
accordingly. However, an ``accepted'' fit (flag 1) does not imply a detection;
where both value and error are zero it means the fitting routine did not
return a solution for absorption that resembled the target spectral feature.

For the sightlines in common between the UVES, Medusa and ARGUS spectra, the
equivalent width of the 4428 \AA\ DIB is measured rather consistently (Fig.\
A1). This is true even for most of the Galactic sightlines. The scatter does
reduce for the combined (LMC + Galactic) absorption, confirming the suspicion
we raised above. The remaining scatter suggests that the errorbars on the
individual measurements are probably a little optimistic.

To give an impression of the quality of the data and of the fits to the DIBs,
in Figure A2 we show six examples of both good and poor signal-to-noise Medusa
spectra, around the 4428 and 6614 \AA\ DIBs (for the same sightlines). To show
the success with which the fitting algorithm measures the DIB in the presence
of challenges we chose to show a spectrum of a B-type star with strong O\,{\sc
ii} lines (VFTS\,28), a very noisy spectrum (VFTS\,258) and a noisy and good
signal-to-noise spectrum with spikes (VFTS\,498 and 689, respectively) as well
as a good signal-to-noise spectrum but with very weak DIBs (VFTS\,333); we
also show the spectrum of the star displaying the strongest DIBs in our
sample, VFTS\,232.

%
% TABLE A2
%
\begin{table*}
\caption{Measurements of the equivalent width of the Galactic (G) and LMC (L)
components of the 4428 and 6614 \AA\ DIBs, probed with the Medusa mode of
VLT--FLAMES. The r.m.s.\ noise values and a flag indicating an accepted (1) or
rejected (0) fit are also tabulated. Only one sample line is displayed; the
full table is available electronically from the CDS.}
\begin{tabular}{l|ccccc|ccccc}
\hline\hline
VFTS                               &
\multicolumn{5}{c|}{4428 \AA\ DIB} &
\multicolumn{5}{c}{6614 \AA\ DIB}  \\
                                   &
EW$_{\rm G}$ (\AA)                  &
f$_{\rm G}$                         &
EW$_{\rm L}$ (\AA)                  &
f$_{\rm L}$                         &
r.m.s.                             &
EW$_{\rm G}$ (\AA)                  &
f$_{\rm G}$                         &
EW$_{\rm L}$ (\AA)                  &
f$_{\rm L}$                         &
r.m.s.                             \\
\hline
542                                &
$0.292\pm0.013$                    &
1                                  &
$0.658\pm0.013$                    &
1                                  &
0.0025                             &
$0.017\pm0.001$                    &
1                                  &
$0.020\pm0.001$                    &
1                                  &
0.0032                             \\
\hline
\end{tabular}
\end{table*}

%
% TABLE A3
%
\begin{table}
\caption{Measurements of the equivalent width of the Galactic (G) and LMC (L)
components of the 4428 \AA\ DIB probed with the ARGUS mode of VLT--FLAMES. The
r.m.s.\ noise values and a flag indicating an accepted (1) or rejected (0) fit
are also tabulated. Only one sample line is displayed; the full table is
available electronically from the CDS.}
\begin{tabular}{lccccc}
\hline\hline
VFTS              &
EW$_{\rm G}$ (\AA) &
f$_{\rm G}$        &
EW$_{\rm L}$ (\AA) &
f$_{\rm L}$        &
r.m.s.            \\
\hline
542               &
$0.220\pm0.021$   &
1                 &
$0.716\pm0.021$   &
1                 &
0.0031            \\
\hline
\end{tabular}
\end{table}

%
% TABLE A4
%
\begin{table*}
\caption{Measurements of the equivalent width of the Galactic (G) and LMC (L)
components of the 4428, 5780 and 5797 \AA\ DIBs, probed with the UVES mode of
VLT--FLAMES. The r.m.s.\ noise values and a flag indicating an accepted (1) or
rejected (0) fit are also tabulated. Only one sample line is displayed; the
full table is available electronically from the CDS.}
\scriptsize
\begin{tabular}{l|ccccc|ccccc|ccccc}
\hline\hline
VFTS                               &
\multicolumn{5}{c|}{4428 \AA\ DIB} &
\multicolumn{5}{c|}{5780 \AA\ DIB} &
\multicolumn{5}{c}{5797 \AA\ DIB}  \\
                                   &
EW$_{\rm G}$ (\AA)                  &
f$_{\rm G}$                         &
EW$_{\rm L}$ (\AA)                  &
f$_{\rm L}$                         &
r.m.s.                             &
EW$_{\rm G}$ (\AA)                  &
f$_{\rm G}$                         &
EW$_{\rm L}$ (\AA)                  &
f$_{\rm L}$                         &
r.m.s.                             &
EW$_{\rm G}$ (\AA)                  &
f$_{\rm G}$                         &
EW$_{\rm L}$ (\AA)                  &
f$_{\rm L}$                         &
r.m.s.                             \\
\hline
542                                &
$0.296\pm0.04$\rlap{7}             &
1                                  &
$0.684\pm0.04$\rlap{7}             &
1                                  &
0.0165                             &
$0.075\pm0.00$\rlap{5}             &
1                                  &
$0.022\pm0.00$\rlap{3}             &
1                                  &
0.0266                             &
$0.023\pm0.00$\rlap{3}             &
1                                  &
$0.000\pm0.00$\rlap{0}             &
1                                  &
0.0266                             \\
\hline
\end{tabular}
\end{table*}

\end{document}